\documentclass[12pt]{article}
\usepackage{amsfonts,epsfig,latexsym,amscd,multicol,theorem,amssymb}
\input diagrams.sty

\newcommand{\disju}{{\perp\!\!\!\perp}} 
\newcommand{\R}{{\mathbb{R}}}
\newcommand{\Z}{{\mathbb{Z}}} 

\newcommand{\C}{{\mathbb{C}}}

\newcommand{\CP}{{\mathbb{C}}{\mathbb{P}}}
\newcommand{\RP}{{\mathbb{R}{{P}}}}
\newcommand{\beq}{\begin{equation}} 
\newcommand{\eeq}{\end{equation}}
\newcommand{\bea}{\begin{eqnarray}} 
\newcommand{\eea}{\end{eqnarray}}
\newcommand{\ra}{\rightarrow} 

\newcommand{\hra}{\hookrightarrow} 

\newcommand{\cd}{\partial} 
\newcommand{\wt}{\widetilde}

\newcommand{\im}{{\rm Im}\, }
 
\newcommand{\Sp}{{\rm Sp}}
\newcommand{\Spin}{{\rm Spin}} 
\newcommand{\SU}{{\rm SU}}
\newcommand{\Su}{{\rm SU}}
 
\newcommand{\susp}{{\rm Susp}}
\newcommand{\hur}{{\rm Hur}} 
\newcommand{\tor}{{\rm Tor}}
\newcommand{\cp}{\smallsmile}

\newcommand{\free}[2]{{\mbox{FreeMaps}}(#1,#2)}
\newcommand{\homo}[2]{{\mbox{Hom}}(#1,#2)}
\newcommand{\QS}[1]{{\rm SU}(#1)^{S^3}}

\theoremstyle{plain} \newtheorem{thm}{Theorem}
\newtheorem{lemma}[thm]{Lemma} \newtheorem{prop}[thm]{Proposition}
 
{\theorembodyfont{\rmfamily}

 \newtheorem{red}{Reduction}
}
\newcommand{\news}{\setcounter{equation}{0}}

\begin{document}

\title{Fermionic quantization and configuration spaces for the Skyrme
and  Faddeev-Hopf models} 

\author{Dave Auckly  \thanks{The first
author was partially supported by  NSF grant DMS-0204651.}  \\
Department of Mathematics, \\ Kansas State University, \\ Manhattan,
Kansas 66506, USA \\ \\ 
J. Martin Speight  \thanks{The second author
was partially supported by EPSRC grant GR/R66982/01.}              \\
School of Mathematics, \\ University of Leeds, \\ Leeds LS2 9JT,
England}

\maketitle

\begin{abstract} 
The fundamental group and rational cohomology of the configuration
spaces of the Skyrme and Faddeev-Hopf models are computed. Physical
space is taken to be a compact oriented 3-manifold, either with or
without a marked point representing an end at infinity. For the Skyrme
model, the codomain is any Lie group, while for the Faddeev-Hopf model
it is $S^2$. It is determined when the topology of configuration space
permits fermionic and isospinorial quantization of the solitons of the
model within generalizations of the  frameworks of
Finkelstein-Rubinstein and Sorkin. Fermionic quantization of Skyrmions
is possible only if the target group contains a symplectic or special
unitary factor, while fermionic quantization of Hopfions is always
possible. Geometric interpretations of the results are given.
\end{abstract}

\section{Introduction}
\label{intro}
\news  
The most straightforward procedure for quantizing a Lagrangian
dynamical system with configuration space $Q$ is to specify the
quantum state by a wavefunction $\psi:Q\ra\C$, but many other
procedures are possible \cite{simwoo}. One may take $\psi$ to be a section of a
complex line bundle over $Q$. Clearly we recover the original
quantization if the bundle is trivial. Depending on the topology of
$Q$ there may be many inequivalent line bundles, giving rise to
quantization ambiguity. The set of equivalence classes of line bundles
is  parametrized by $H^2(Q,\Z)$. Quantization ambiguity can be used to
generate fermionic or bosonic quantizations of the same classical
system. The example of a charged particle under the influence of a
magnetic monopole studied by Dirac \cite{dirac1, dirac2} clearly
demonstrates the utility of complex line bundles for analyzing quantum
dynamics. See also, the discussion in \cite{z} and \cite[chapter
6]{bmss}.

In the case of a Lagrangian field theory supporting topological
solitons, configuration space is typically the space of (sufficiently
regular) maps from some 3-manifold (representing physical space) to
some target manifold. A famous example of this is the Skyrme model with target
space $\SU(N)$, $N\geq 2$ and physical space $\R^3$.  Here fermionic
quantization is phenomenologically crucial since the solitons are
taken to represent  protons and
neutrons. 

Recall the distinction between bosons and fermions: a macroscopic ensemble
 of identical bosons behaves statistically as if
aribtrarily many particles can lie in the same state, while a
macroscopic ensemble of identical fermions behaves as if no two particles can
lie in the same state. Photons are examples of particles with
bosonic statistics and electrons are examples of particles with
fermionic statistics. There are several theoretical models of particle
statistics. In quantum mechanics, the wavefunction representing a
multiparticle state is symmetric under exchange of any pair of
identical bosons, and antisymmetric under exchange of any pair of
identical fermions.  In conventional perturbative quantum field
theory, commuting fields are used to represent bosons and
anti-commuting fields are  used to represent fermions. More precisely,
bosons have commuting creation operators and fermions have
anti-commuting creation operators. However, there are times when
fermions may arise within a field theory  with purely bosonic
fundamental fields. This phenomenon is called emergent fermionicity,
and it relies crucially on the topological properties of the
underlying configuration space of the model. Analogous instances of
emergent fermionicity in quantum mechanical (rather than field
theoretic) settings are described in \cite{B, sch,z} and \cite[chapter
7]{bmss}. 

When Skyrme originally proposed
his model, it was not clear how fermionic quantization of the solitons
could be achieved, a fundamental gap which he acknowledged
\cite{sky}. 
The possibility of fermionically quantizing the Skyrme model was first 
demonstrated (for $N=2$) by
Finkelstein and Rubinstein \cite{finrub}. Full consistency of their
quantization procedure was established by Giulini \cite{giu}.  The
case $N\geq 3$ was dealt with in a rather different way by Witten
\cite{wit} at the cost of introducing a new term into the Skyrme
action.  This was a crucial development, since the $N=3$ model is
particularly phenomenologically favoured. Although the approaches of
Finkelstein-Rubinstein and Witten appear quite different, they can be
treated in a common  framework, as demonstrated by Sorkin
\cite{sorkin}. See \cite{at1} and  \cite{bmss} for exposition about
where the Skyrme model fits into modern  physics. Also see \cite{at2},
and \cite{ramadas} for a discussion of  fermionic quantization of
$\SU(N)$ valued Skyrme fields. We will review the Finkelstein-Rubinstein and 
Sorkin models of particle statistics in section \ref{phy} below, and 
describe the obvious generalization of their models when the domain of the 
soliton is something other than $\R^3$.

Spin is a property of a particle state associated with how it transforms
under spatial rotations. Let us briefly
review the usual mathematical models of spin. There are two
general situations. In one, space admits a global rotational
symmetry, in the other, it does not.  
When physical space is $\R^3$, so space-time is usual
Minkowski space, the classical rotational symmetries induce quantum
symmetries that are representations of the $Spin$ group. These
irreducible representations are labeled by half integers (one half of
the number of boxes in a Young diagram for a representation of
SU$(2)$.)  The integral representations are honest representations
of the rotation group, but the fractional ones are not. The spin of a 
particle is the half integer labelling the $\Su(2)$ representation under
which its wavefunction transforms. If the spin is not an integer, the
particle
is said to be spinorial.
When physical
space is not $\R^3$, one can instead consider the bundle of frames
over physical space (vierbeins). Spin may then be modeled by the
action of the rotation group on these frames \cite{birrell}.  The
easiest case in this direction is when space-time admits a spin
structure. This reconciliation of  spin with the possibility of a
curved space time was an important discovery in the last century. The
situation is parallel in nonlinear models. In the easiest example of
of a model with solitons incorporating spin, the configuration space
is not $\R^3$, but maps defined on $\R^3$. The rotation group acts on
such maps by precomposition. Sorkin described a model for the spin of
solitons which generalizes to any maps defined on any domain. We cover
this in section \ref{phy}.  

Although spin and particle statistics have completely different conceptual
origins, there are strong connexions between the two.
The spin-statistics
theorem asserts, in the context of axiomatic quantum field theory,
that particles are fermionic if and only if they are spinorial. Said
differently, any particle with fractional spin is a fermion, and any
particle with integral spin is a boson. Analogous spin-statistics 
theorems have been found for solitons also.
Such a theorem was proved for $\SU(2)$ Skyrmions on $\R^3$ by
Finkelstein and Rubinstein \cite{finrub}, and for arbitrary textures
on $\R^3$ by Sorkin \cite{sorkin}, using only the topology of
configuration space. By a texture, we mean that the
field must approach a constant limiting value at spatial infinity, in
contrast to, say, monopoles and vortices. So pervasive is the link between fermionicity and
spinoriality that the two are often elided. For example,
Ramadas argued that it was
possible to fermionically quantize SU$(N)$ Skyrmions on $\R^3$ because
it was possible to spinorially quantize them \cite{ramadas}.

Isospin is the conserved quantity analogous to spin associated with
 an {\em internal} rotational 
symmetry. As in the simplest model of spin, a particle's 
isospin is a half integer labelling the representation of the quantum
symmetry group corresponding to the classical internal ${\rm SO}(3)$
symmetry.
In all the models we consider, the target space
has a natural $SO(3)$ action, so it will make sense to determine
whether these models admit isospinorial quantization in the 
usual sense.
Krusch  \cite{kru} has shown that $\SU(2)$
Skyrmions are spinorial if and only if they are isospinorial, which is
in good agreement with nuclear phenomenology, since they represent
bound states of nucleons.  Recall that 
nucleon is the collective term for the proton
and neutron. Both have spin and isospin 1/2 but are in  different
eigenstates of the 3rd component of isospin: the proton has $I_3=1/2$,
and the neutron has $I_3=-1/2$. In general, the integrality of spin is
unrelated to that of isospin. Strange hadrons can have half-integer
spin and integer isospin (and vice-versa). For example, the
$\Sigma$-baryon has isospin $1$, but spin $1/2$, and the $K$
mesons  have isospin $1/2$ and spin $1$. One would hope, therefore,
that the correlation found by Krusch fails in the $\Su(N)$ Skyrme
model with $N>2$, since this is supposed to model low energy QCD with
more than two light quark flavours, and should therefore be able to
accommodate the more exotic baryons. The mathematical reader
unfamiliar with spin, isospin, strangeness etc.\ may find the book by
Halzen and Martin \cite{HM} and the comprehensive listing of particles
and their properties in \cite{pdg} helpful.

Emergent fermionicity, like (iso)spinoriality, 
can often be incorporated into a quantum system
by exploiting the possibility of 
differences between the classical and quantum symmetries of
 the space of quantum states \cite[chapter 7]{bmss}. A spinning top is
 a well known example of this. The classical symmetry group is
 SO$(3)$, while the quantum symmetry group for some quantizations is
 SU$(2)$, \cite{B, sch}. An electron in the field of a magnetic
 monopole is also a good example, \cite[chapter 7]{bmss} and
 \cite{z}. We emphasize, however, that 
emergent fermionicity does {\em not} depend on any symmetry assumptions. In
 fact, the model of particle statistics that we mainly consider
 (Sorkin's model) depends only on the topology of the configuration
 space.

The purpose of this paper is to determine the quantization ambiguity
for a wide class of field theories supporting topological solitons of
texture type in 3 spatial dimensions. We will allow (the one
point compactification of) physical space to be any compact, oriented
3-manifold $M$ and target space to be any Lie group $G$, or the
2-sphere. The results use only the topology of configuration space and
are completely independent of the dynamical details of the field
theory. They cover, therefore, the Faddev-Hopf and general Skyrme
models on any orientable domain.  Our main mathematical results will
be the computation of the fundamental group and the rational or real
cohomology ring of $Q$. (The universal coefficient theorem implies
that the rational dimension of the rational cohomology is equal to the
real dimension of the real cohomology. Homotopy theorists tend to
express results using rational coefficients and physicists tend to use
real or complex coefficients.)  We shall see that quantization ambiguity, as 
described by $H^2(Q,\Z)$, may be reconstructed from these
data. We also give geometric interpretations of the algebraic results
which are useful for purposes of visualization. We then determine
under what circumstances the quantization ambiguity allows for
consistent fermionic quantization of Skyrmions and hopfions within the
frameworks of Finkelstein-Rubinstein and Sorkin. We finally discuss the 
spinorial and isospinorial quantization of these models.

The main motivation for this work was to test the phenomenon of
emergent fermionicity (i.e.\ fermionic solitons in a bosonic theory)
in the Skyrme model to see, in particular, whether it survives the
generalization from domain $\R^3$ to domain $M$. Our philosophy is
that a concept in field theory which cannot be properly formulated on
any oriented domain should not be considered fundamental. In fact, we
shall see indications that emergent  fermionicity is insensitive to
the topology of $M$, but depends crucially on  the topology of the
target space. 

We would like to thank Louis Crane, Steffen Krusch and Larry Weaver for 
helpful conversations about particle  physics.

\section{Notation and statement of results}
\label{not}
\news

Recall that topologically distinct complex line bundles over a
topological  space $Q$ are classified by $H^2(Q,\Z)$. Note that $Q$
need have no differentiable structure to make sense of this: we can
define $c_1(L) \in H^2(Q,\Z)$ corresponding to bundle $L$ directly in
terms of the transition functions of $L$ rather than thinking of it as
the curvature of a unitary connexion on $L$ \cite{simwoo}. So this
classification applies in the cases of interest. The free part of
$H^2(Q,\Z)$ is determined by $H^2(Q,\R)$, while its torsion is
isomorphic to the torsion of $H_1(Q)$. For any topological space,
$H_1(Q)$ is isomorphic to the abelianization of $\pi_1(Q)$.  The
bundle classification problem is solved, therefore, once we know
$\pi_1(Q)$ and $H^2(Q,\R)$.

In this section we will define the configuration spaces that we
consider,  set up notation and state our main topological
results. There are in fact  many different but related configuration
spaces that we could consider (for  example spaces of free maps versus
spaces of base pointed maps) and several  different possibilities
depending on whether the domain is connected etc.  We give clean
statements of our results for special cases in this section,  and
describe how to obtain the most general results in the next section.
The next section will also include some specific examples. Of course
homotopy theorists have studied the algebraic topology of spaces of
maps.  The paper by Federer gives a spectral sequence whose limit
group is a sum  of composition factors of homotopy groups for a space
of based maps  \cite{fed}. We do not need a way to compute -- we
actually need the  computations, and this is what is contained here.

Let $M$ be a compact,  oriented 3-manifold and $G$ be any Lie group.
Then the first configuration space we consider is either
$\free{M}{G}$, the space of continuous maps $M\ra G$, or $G^M$, the
subset of $\free{M}{G}$ consisting of those maps which send a chosen
basepoint $x_0\in M$ to $1\in G$. We will address configuration spaces
of  $S^2$-valued maps later in this section and paper. Both
$\free{M}{G}$ and  $G^M$ are given the compact open topology.  In
practice some Sobolev topology depending on the energy functional is
probably appropriate. The issue of checking the algebraic topology
arguments given in this paper for classes of Sobolev maps is
interesting.  See \cite{AK3} for a discussion of the correct setting
and arguments  generalizing the labels of the path components of these
configuration spaces  for Sobolev maps. The space $\free{M}{G}$ is
appropriate to the $G$-valued  Skyrme model on a genuinely compact
domain, while $G^M$ is appropriate to the case where physical space
$\hat{M}$ is noncompact but has a connected end at infinity which, for
finite energy maps, may be regarded as a  single point $x_0$ in the
one point compactification $M$ of $\hat{M}$. 

The space $G^M$ splits into disjoint path components which are labeled
by  certain cohomology classes on $M$ \cite{AK1}. Let $(G^M)_0$ be the
identity  component of the topological group $G^M$, that is, the path
component containing the constant map $u(x)=1$. In physical terms
$(G^M)_0$ is the vacuum sector of  the model. Then $(G^M)_0$ is a
normal subgroup of $G^M$ whose cosets are precisely the other path
components. The set of path components itself has a natural group
structure. As a set, the space of path components of the based maps is
given by the following proposition.

\begin{prop}[Auckly-Kapitanski]\label{components}
Let $G$ be a compact, connected Lie group and $M$ be a connected,
closed $3$-manifold.  The set of path components of $G^M$ is
$$
G^M/(G^M)_0\,\cong\, H^3(M; \pi_3(G))\times H^1(M; H_1(G)).
$$
\end{prop} 

\noindent
The reason the above proposition only describes the set of path
components is that Auckly and Kapitanski only establish an exact
sequence
$$
0\to H^3(M; \pi_3(G))\to G^M/(G^M)_0\to H^1(M; H_1(G))\to 0.
$$
To understand the group structure on the set of path components one
would have to understand a bit more about this sequence (e.g. does it
split?).  Every path component of $G^M$ is homeomorphic to $(G^M)_0$
since $\wt{u}(x)\mapsto u(x)^{-1}\wt{u}(x)$ is a homeomorphism $u\,
(G^M)_0\ra (G^M)_0$.  Our first result computes  the fundamental group
of the configuration space of based $G$-valued maps.

\begin{thm}\label{thm1} If $M$ is a closed, connected, orientable 
$3$-manifold, and $G$ is any Lie group, then 
$$
\pi_1(G^M)\, \cong\,{\mathbb Z}_2^s\,\oplus H^2(M; \pi_3(G)).
$$
Here $s$ is the number of symplectic factors in the lie algebra of $G$.
\end{thm}

\noindent
Our next result gives the whole real cohomology ring
$H^*((G^M)_0,\R)$, including its multiplicative structure. This, of
course, includes the required computation of $H^2((G^M)_0,\R)$.

Similarly to Yang-Mills theory, there is a $\mu$ map, 
$$
\mu:H_d(M;{\mathbb R})\otimes H^j(G;{\mathbb R}) \to
H^{j-d}(G^M;{\mathbb R}),
$$
and the cohomology ring is generated (as an algebra) by the images of
this  map. To state the theorem we do not need the definition of this
$\mu$ map,  but the definition may be found in subsection
\ref{cohomdesc} of section  \ref{geo}, in particular, equation
(\ref{mudef}).

\begin{thm}\label{thm1co} Let $G$ be a compact, simply-connected, simple Lie 
group. The cohomology ring of any of these groups is a free
graded-commutative unital algebra over $\R$ generated by degree $k$
elements $x_k$ for certain values of $k$ (and with at most one
exception at  most one generator for any given degree). The values of
$k$ depend on the  group and are listed in table \ref{tbl1} in section
\ref{hom}. Let $M$ be  a closed, connected, orientable
$3$-manifold. The cohomology ring  $H^*((G^M)_0;\R)$ is the free
graded-commutative unital algebra over $\R$ generated by the elements
$\mu(\Sigma_j^d\otimes x_k)$,  where $\Sigma_j^d$ form a basis for
$H_d(M;\R)$ for $d>0$ and $k-d>0$. 
\end{thm}
\noindent
The examples in the next section best illuminate the details of the
above  theorem.

Turning to the Faddeev-Hopf model, the configuration space of interest
is either the space of free $S^2$-valued maps $\free{M}{S^2}$, or
$(S^2)^M$, the space of based continuous maps $M\ra S^2$. One can
analyze $\free{M}{S^2}$ in terms of $(S^2)^M$ by making use of the
natural fibration  
$$ (S^2)^M\hookrightarrow
\free{M}{S^2}\stackrel{\pi}{\ra}S^2,\qquad \pi:u(x)\mapsto u(x_0).
$$
 The fundamental cohomology class (orientation class),
$\mu_{S^2}\in H^2(S^2,\Z)$ plays an important role in the description
of the mapping spaces of $S^2$-valued maps.  The path components of
$(S^2)^M$ were determined by Pontrjagin \cite{pont}:

\begin{thm}[Pontrjagin] 
\label{ponthm}
Let $M$ be a closed, connected, oriented 3-manifold, and $\mu_{S^2}$
be a generator of $H^2(S^2;\Z)\cong\Z$.  To any based map $\varphi$
from $M$ to $S^2$ one may associate the cohomology  class,
$\varphi^*\mu_{S^2}\in H^2(M;\Z)$. Every second cohomology class may
be obtained from some map and any two maps with different cohomology
classes  lie in distinct path components of $(S^2)^M$. Furthermore,
the set of path  components corresponding to a cohomology class,
$\alpha\in H^2(M)$ is in  bijective correspondence with
$H^3(M)/(2\alpha\cp H^1(M))$.
\end{thm}

\noindent
A discussion of this theorem in the setting of the Faddeev model may
be found in  \cite{AK2} and \cite{AK3}.  Let $(S^2)^M_0$ denote the
vacuum sector, that is the path component of the constant map, and
$(S^2)^M_\varphi$ denote the path component containing
$\varphi$. Since $(S^2)^M$ is not a topological group, there is no
reason to expect all its path components to be homeomorphic. We will
prove, however that two components $(S^2)^M_\varphi$ and
$(S^2)^M_\psi$ are homeomorphic if $\varphi^*\mu_{S^2} =
\psi^*\mu_{S^2}$:

\begin{thm}
\label{fhhom}   
Let $\varphi\in (S^2)^M$ such that $\varphi^*\mu_{S^2}=
\psi^*\mu_{S^2}$. Then  $(S^2)^M_\varphi\cong(S^2)^M_\psi$.
\end{thm}

\noindent
Moreover, the fundamental group of any component can be computed, as
follows.

\begin{thm}\label{thm2}  Let $M$ be closed, connected and orientable. For 
any $\varphi\in (S^2)^M$,
the fundamental group of $(S^2)^M_\varphi$ is given by
$$
\pi_1((S^2)^M_\varphi)\cong {\mathbb Z}_2\oplus H^2(M;{\mathbb
Z})\oplus  \hbox{\rm ker}(2\varphi^*\mu_{S^2}\cp).
$$
Here $2\varphi^*\mu_{S^2}\cp:H^1(M;{\mathbb Z})\to  H^3(M;{\mathbb
Z})$ is the usual map given by the cup product. 
\end{thm}

There is a general relationship between the fundamental group of the
configuration space of based $S^2$-valued maps and the corresponding
configuration space of free maps. It implies the following result for
the fundamental group of the space of free $S^2$-valued maps.

\begin{thm}\label{freefun}
We have $\pi_1(\free{M}{S^2}_\varphi)\cong {\mathbb Z}_2 \oplus
\left(H^2(M;{\mathbb Z})/ \langle 2\varphi\mu_{S^2}\rangle
\right)\oplus  \hbox{\rm ker}(2\varphi^*\mu_{S^2}\cp)$. 
\end{thm}

To complete the classification of complex line  bundles over
$(S^2)^M_\varphi$ one also needs the second cohomology  $H^2((S^2)^M_\varphi;\R)$, which can
be extracted from the above theorem and the following computation of
the cohomology ring  $H^*((S^2)^M_\varphi,\R)$:

\begin{thm}\label{thm2co}
Let $M$ be closed, connected and orientable, let $\varphi:M\to S^2$,
let $\Sigma_j^d$ form a basis for $H_d(M;\R)$ for $d<3$, and let
$\{\alpha_k\}$ for a basis for
$\hbox{ker}(2\varphi^*\mu_{S^2}\cup):H^1(M;\Z)\to H^3(M;\Z))$. The
cohomology ring $H^*((S^2)^M_\varphi;\R)$ is the free
graded-commutative unital algebra over $\R$ generated by the elements
$\alpha_k$ and $\mu(\Sigma_j^d\otimes x)$,  where $x\in H^3(Sp_1;\Z)$
is the orientation class. The classes $\alpha_k$ have degree $1$ and
$\mu(\Sigma_j^d\otimes x)$ have degree $3-d$.
\end{thm}

We can compute the cohomology of the space of free $S^2$-valued maps
using the following theorem.
\begin{thm}\label{freeco}
There is a spectral sequence with $E_2^{p,q}=H^p(S^2;\R)\otimes
H^q((S^2)^M_\varphi;\R)$ converging to
$H^*(\free{M}{S^2}_\varphi;\R)$. The second differential is given by
$d_2\mu(\Sigma^{(2)}\otimes x)=2\varphi^*\mu_{S^2}[\Sigma]\mu_{S^2}$
with $d_2$ of any other generator trivial. All higher differentials
are trivial as well.
\end{thm}

In order to compare the classical and quantum isospin symmetries, we
will  use the following theorem due to Gottlieb \cite{Got}. It is
based on  earlier work of Hattori and Yoshida \cite{HY}.

\begin{thm}[Gottlieb]\label{thmgot}
Let $L\to X$ be a complex line bundle over a locally compact space. An
action of a compact connected lie group on $X$, say $\rho:X\times G\to
X$,  lifts to a bundle action on $L$ if and only if two obstructions
vanish. The  first obstruction is the pull back of the first chern
class,  $L_{x_0}^*c_1(L)\in H^2(G;\Z)$. Here $L_{x_0}$ is the map
induced by  applying the group action to the base point. The second
obstruction lives  in $H^1(X;H^1(G;\Z))$.
\end{thm}

\noindent We have taken the liberty of radically changing the notation
from  the original theorem, and we have only stated the result for
line bundles.  The actual theorem is stated for principal torus
bundles. Since our  configuration spaces are not locally compact, we
should point out that  we will use one direction of this theorem by
restricting to a locally  compact equivariant subset. In the other
direction, we will just outline a  construction of the lifted action.

Our main physical conclusions are:
\begin{description}
\item[C1]In these models, there is a portion of
quantization  ambiguity that depends only on the codomain and is
completely independent  of the topology of the domain. This allows for
the possibility that  emergent fermionicity may only depend on the
target.
\item[C2] It is possible to quantize $G$-valued
solitons fermionically (with odd exhange statistics) if and only if the Lie algebra 
contains a
symplectic  ($C_n$) or special unitary ($A_n$) factor.
\item[C3] It is possible to quantize $G$-valued solitons
with  fractional isospin when the Lie algebra of $G$ contains a
symplectic  ($C_n$) or special unitary ($A_n$) factor.
\item[C4] It is not possible to quantize $G$-valued
solitons  with fractional isospin when the Lie algebra does not
contain such a  factor.
\item[C5] It is always possible to choose a quantization
of  these systems with integral isospin (however such might not be
consistent  with other constraints on the model)
\item[C6] It is always possible to quantize $S^2$-valued
solitons with fractional isospin and odd exchange statistics.
\end{description}

The rest of this paper is structured as follows. In section \ref{hom}
we describe how to reduce the description of the topology of general
$G$-valued and $S^2$-valued mapping spaces to the theorems listed  in
this section. We also provide several illustrative examples.  In
section \ref{geo} we review the Pontrjagin-Thom construction and
describe geometric interpretations of some of our results using the
Pontrjagin-Thom construction. Physical applications, particularly the
possibility of consistent fermionic quantization of Skyrmions, are
discussed in section \ref{phy}. Finally, section \ref{pro} contains
the proofs of our results.

\section{Preliminary reductions and examples}
\label{hom}\label{reduct}
\news

We begin this section with a collection of observations that allow one
 to reduce questions about the topology of  various mapping spaces of
 $G$-valued and $S^2$-valued maps to the theorems  listed in the
 previous section. Many of these observations will reduce a  more
 general mapping space to a product of special mapping spaces, or put
 such spaces into fibrations. These reductions ensure that our results
 are  valid for arbitrary closed, orientable $3$-manifolds, and valid
 for  {\it any} Lie group.

It follows directly from the definition of $\pi_1$ that
$\pi_1(X\times Y) \cong \pi_1(X)\times\pi_1(Y)$. The cohomology of a
product is described by the K\"unneth theorem, see \cite{span}. For
real  coefficients it takes the simple form, $H^*(X\times Y)\cong
H^*(X)\otimes  H^*(Y)$. The cohomology ring of a disjoint union of
spaces is the direct  sum of the corresponding cohomology rings,
i. e. $H^*(\disju X_\nu;A)= \bigoplus H^*(X_\nu;A)$. Recall that a
fibration is a map with the covering  homotopy property, see  for
example \cite{span}. Given a fibration $F\hra E\ra B$, there is an
induced long exact sequence of homotopy groups,  $\dots\ra
\pi_{k+1}(B)\ra \pi_k(F)\ra \pi_k(E)\ra\pi_k(B)\ra \dots$,  see
\cite{span}. By itself this sequence is not enough to determine the
fundamental group of a term in a fibration from the other
terms. However,  combined with a bit of information about the twisting
in the bundle it will  be enough information. One can also relate the
cohomology rings of the  terms in a fibration. This is accomplished by
the Serre spectral sequence,  see \cite{span}.

\begin{red}\label{r1}
We have, $\free{ \disju X_\nu}{Y}\cong \prod\free{X_\nu}{Y}$ and
$Y^{\disju X_\nu}\cong Y^{X_0}\times \prod_{\nu\ne 0}\free{X_\nu}{Y}$
where  $X_0$ is the component of $X$ containing the base point.
\end{red}

It follows that there is no loss of generality in assuming that $M$ is
connected. Likewise there is no loss of generality in assuming that
the  target is connected because of the following reduction.

\begin{red}\label{r2}
We have, $\free{X}{\disju Y_\nu}\cong\disju\free{X}{Y_\nu}$ and
assuming  $X$ is connected $Y^X=Y_0^X$ where $Y_0$ is the component
containing the  base point.
\end{red}

Both $\free{M}{G}$ and $G^M$ are topological groups under pointwise
multiplication.  In fact $\free{M}{G}\cong G^M\rtimes G$, the
isomorphism being $u(x)\mapsto (u(x)u(x_0)^{-1},u(x_0))$, which is
clearly a homeomorphism $\free{M}{G}\ra G^M\times G.$ It is thus
straightforward to deduce  $\pi_1(\free{M}{G}$ and
$H^*(\free{M}{G},\R)$ from $\pi_1(G^M)$ and  $H^*(G^M,\Z)$. Note that
the based case includes the standard choice  $\hat{M}=\R^3$.

\begin{red}\label{r3}
We have $\free{M}{G}\cong G^M\times G$.
\end{red}

In the same way, we can reduce the free maps case to the based case
for  $S^2$-valued maps. In this case we only obtain a fibration. See
Lemmas  \ref{s2tofree}, \ref{Fhseq} and \ref{Fhsplit}.

\begin{red}\label{r4}
We have a fibration, $(S^2)^M\hra \free{M}{S^2}\ra S^2$,
$\pi_0(\free{M}{S^2})=\pi_0((S^2)^M)$, and
$\pi_1(\free{M}{S^2}_\varphi)= \pi_1((S^2)^M_\varphi)$.
\end{red}

\noindent
The relevant information about the twisting in this fibration as far
as the  fundamental group detects is given in the proof of Theorem
\ref{freefun}  contained in subsection \ref{subsecpc}. For cohomology,
the information is  encoded in the second differential of the
associated spectral sequence.  Returning to the case of group-valued
maps we know by the  Cartan-Malcev-Iwasawa theorem that any connected
Lie group is homeomorphic  to a product $G=K\times \R^n$, where $K$ is
compact \cite{iwa}.

\begin{red}\label{r5}
If $X^\prime$ and $Y^\prime$ are homotopy equivalent to $X$ and $Y$
respectively, then $(Y^\prime)^{X^\prime}$ is homotopy equivalent to
$Y^X$.  In particular we have $G^M\simeq K^M$.
\end{red}

Recall that every path component of $G^M$ is homeomorphic to $(G^M)_0$
We may  therefore consider only the vacuum sector $(G^M)_0$, without
loss of  generality. We shall see that things are very different for
the Faddeev-Hopf configuration  space, where we must keep track of
which path component we are studying.

\begin{red}\label{r6}
If $G$ is a Lie group, $\tilde{G}$ is the universal covering group of
its identity component and $M$ is a $3$-manifold, then
$$(\tilde G^M)_0 \cong (G^M)_0.
$$
\end{red}
\medskip
{\it Proof:} Without loss of generality we may assume that $G$ is
connected.  We have the exact sequence,
$$
1\to\tilde G^M\to G^M \to H^1(M;H_1(G))\to 0,
$$
from \cite{AK1}. The exactness follows from the unique path lifting
property  of covers at the first term, the lifting criteria for maps
to the  universal  cover at the center term, and induction on the
skeleton of $M$ at  the last term. Clearly, the identity component of
$\tilde G^M$ maps to the  identity component of $G^M$. By the above
sequence, this map is injective.  Any element of $(G^M)_0$, say $u$,
maps to $0$ in $H^1(M;H_1(G))$, so is  the  image of some map in
$\tilde G^M$, say $\tilde u$. Using the  homotopy lifting property of
covering spaces, we may lift  the homotopy of $u$ to a constant map,
to a homotopy of $\tilde u$ to a  constant map and conclude that
$\tilde u\in(\tilde G^M)_0$. It follows that  the map,  $(\tilde
G^M)_0\to (G^M)_0$ is a homeomorphism. \hfill $\Box$

\begin{red}\label{r7}
The universal covering group of any compact Lie group is  a product of
${\mathbb R}^m$  with a finite number of compact, simple,
simply-connected factors  \cite{Mimura}. Furthermore, $\left(\prod
Y_\nu\right)^X \cong \prod  Y_\nu^X$ and $\free{X}{\prod
Y_\nu)}\cong\prod\free{X}{Y_\nu}$. 
\end{red}

We have therefore reduced to the case of closed, connected, orientable
$M$  and compact, simple, simply-connected Lie groups.  All compact,
simple, simply-connected Lie groups are listed together with their
center  and rational cohomology in table  \ref{tab1}.


\begin{table}\label{tbl1}\label{table1}
\centering
\begin{tabular}{|l|l|l|}
\hline
group, $G$ & center, $Z(G)$ &
$H^*(G;{\mathbb Q})$ \\ \hline $A_n= \hbox{SU}(n+1)$, $n\ge 2$ &
${\mathbb Z}_{n+1}$  &  ${\mathbb Q}[x_3, x_5, \dots x_{2n+1}]$ \\
\hline $B_n=\hbox{Spin}(2n+1)$, $n\ge 3$ & ${\mathbb Z}_2$ &
${\mathbb Q}[x_3, x_7, \dots x_{4n-1}]$ \\ \hline $C_n=\hbox{Sp}(n)$,
$n \ge 1$ & ${\mathbb Z}_2$ &  ${\mathbb Q}[x_3, x_7, \dots x_{4n-1}]$
\\ \hline $D_n=\hbox{Spin}(2n)$, $n \ge 4$ &  ${\mathbb Z}_2\oplus
{\mathbb Z}_2$ for $n\equiv_2 0$ &  ${\mathbb Q}[x_3,  x_7, \dots
x_{4n-5}, y_{2n-1}]$  \\ \, & ${\mathbb Z}_4$ for $n\equiv_2 1$ & \\
\hline $E_6$ & ${\mathbb Z}_3$ &  ${\mathbb Q}[x_3, x_9, x_{11},
x_{15}, x_{17}, x_{23}]$ \\ \hline $E_7$ & ${\mathbb Z}_2$ &
${\mathbb Q}[x_3, x_{11}, x_{15}, x_{19}, x_{23}, x_{27}, x_{35}]$ \\
\hline $E_8$ & 0 &  ${\mathbb Q}[x_3, x_{15}, x_{23}, x_{27}, x_{35},
x_{39}, x_{47}, x_{59}]$ \\ \hline $F_4$ & 0 &  ${\mathbb Q}[x_3,
x_{11}, x_{15}, x_{23}]$ \\ \hline  $G_2$ & 0 &  ${\mathbb Q}[x_3,
x_{11}]$ \\  \hline%
\end{tabular}
\caption{Simple groups}\label{tab1}  
\end{table}

Recall from Proposition \ref{components} that the path components of a
configuration space of group-valued maps depend on the fundamental
group  of the group.  The fundamental group of any Lie group is a
discrete subgroup of the  center  of the universal covering group.
The center of such a group is just the product of the centers of the
factors.

Some comments about table \ref{tab1} are in order at this point. The
last  generator of the cohomology ring of $D_n$ is labeled with a $y$
instead of  an $x$ because there are two generators in degree $2n-1$
when $n$ is even.  As usual,  $\hbox{SU}(k)$ is the set of  special
unitary matrices, that is complex matrices with unit determinant
satisfying, $A^*A=I$. The symplectic groups,  $\hbox{Sp}(k)$, consist
of the quaternionic  matrices satisfying $A^*A=I$, and the special
orthogonal groups,  $\hbox{SO}(k)$ consist of the real matrices  with
unit determinant satisfying $A^*A=I$.  The spin groups,
$\hbox{Spin}(k)$ are the universal covering groups of the special
orthogonal groups. The definitions of the exceptional groups may be
found in \cite{Adams}. The following  isomorphisms hold,
$\hbox{SU}(2)\cong\hbox{Sp}(1)\cong\hbox{Spin}(3)$,
$\hbox{Spin}(5)\cong\hbox{Sp}(2)$, and
$\hbox{Spin}(6)\cong\hbox{SU}(4)$,  \cite{Adams}. We will need some
homotopy groups of Lie groups. Recall that the higher homotopy groups
of a space are isomorphic to the higher homotopy  groups of the
universal cover of  the space, and the higher homotopy groups take
products to products.  We  have $\pi_3(G)\cong {\mathbb Z}$ for  any
of the simple $G$, and $\pi_4(\hbox{Sp}(n))\cong {\mathbb Z}_2$ and
$\pi_4(G)=0$ for all other simple groups \cite{Mimura}. This is the
reason we grouped the simple  groups  as we did. Note in particular
that we are calling SU$(2)$ a symplectic group.

\subsection{Examples}\label{ex}

In this subsection, we present two examples that suffice to illustrate
all  seven reductions described earlier.

\noindent
{\bf Example 1} For our first example, we take $M=(S^2\times
S^1)\disju  \RP^3$ and 
$$
G=\hbox{Sp}(2)\times\left\{\left(\begin{array}{cc} a
&b\\0&c\end{array} \right)\in\hbox{GL}_2\R\right\}.
$$
We take $((1,0,0),(1,0))\in S^2\times S^1$ as the base point in
$M$. In  this example, neither the domain nor codomain is connected
($G/G_0\cong  \Z_2\times\Z_2$). In addition, the group is not
reductive. We also see  exactly what is meant by the number of
symplectic factors in the Lie  algebra: it is just the number of $C_n$
factors in the Lie algebra of the maximal compact subgroup of the
identity  component of $G$. This example requires Reductions \ref{r1},
\ref{r2},  \ref{r3}, and \ref{r5}.  To analyze the topology of the
spaces of free and based maps, it suffices  to understand maps from
$S^2\times S^1$ and $\RP^2$ into the identity  component, $G_0$
(Reductions \ref{r1}, \ref{r2} and \ref{r3}). 
In fact, we may replace $G_0$  with
$\hbox{Sp}(2)$ (Reduction \ref{r5}). Proposition \ref{components} implies
that $\pi_0(\hbox{Sp}(2)^{S^2\times S^1})=\Z$ and
$\pi_0(\hbox{Sp}(2)^{\RP^3})=\Z$, so $\pi_0(\free{M}{G})=\Z_2^4\times
\Z^2$  and $\pi_0(G^M)=\Z_2^2\times\Z^2$.  Similarly, Theorem
\ref{thm1} implies that $\pi_1(\hbox{Sp}(2)^{S^2\times
S^1})=\Z_2\oplus\Z$ and $\pi_1(\hbox{Sp}(2)^{\RP^3})=\Z_2\oplus\Z_2$,
so  $\pi_1(\free{M}{G})=\pi_1(G^M)=\Z_2^3\oplus \Z$. 

Turning to the cohomology, we know that $H^*(\hbox{Sp}(2);\R)$ is the
free  graded-commutative unital algebra generated by $x_3$ and
$x_7$. Graded-commutative means $xy=(-1)^{|x||y|}yx$. It follows that
any term with repeated factors is zero. We can list the generators of
the groups in each degree. In the expression below we list the
generators left to right from degree $0$ with each degree separated by
vertical lines:
$$
H^*(\hbox{Sp}(2);\R)=|1|0|0|x_3|0|0|0|x_7|0|0|x_3x_7|.
$$
The product structure is apparent. Theorem \ref{thm1co} tells us that
$H^*((\hbox{Sp}(2))^{S^2\times S^1}_0;\R)$ is the free unital
graded-commutative algebra generated by $\mu([S^2\times
\hbox{pt}]\otimes x_3)^{(1)}$, $\mu([\hbox{pt}\times S^1]\otimes
x_3)^{(2)}$, $\mu([S^2\times S^1]\otimes x_7)^{(4)}$,
$\mu([S^2\times \hbox{pt}]\otimes x_7)^{(5)}$, and
$\mu([\hbox{pt}\times S^1]\otimes x_7)^{(6)}$. Here we have included
the degree of the generator as a subscript. In the same way we see
that $H^*((\hbox{Sp}(2))^{\RP^3}_0;\R)$ is the free unital
graded-commutative algebra (FUGCA) generated by $\mu([\RP^3]\otimes
x_7)^{(4)}$. Using the reductions and the K\"unneth theorem we see
that $H^*((G^M)_0;\R)$ is the FUGCA generated by $\mu([S^2\times
\hbox{pt}]\otimes x_3)^{(1)}$, $\mu([\hbox{pt}\times S^1]\otimes
x_3)^{(2)}$, $x_3$, $\mu([S^2\times S^1]\otimes x_7)^{(4)}$,
$\mu([\RP^3]\otimes x_7)^{(4)}$,   $\mu([S^2\times \hbox{pt}]\otimes
x_7)^{(5)}$, $\mu([\hbox{pt}\times S^1]\otimes x_7)^{(6)}$, $x_7$.
Notice that this is not finitely generated as a vector space even
though it is finitely generated as an algebra. This is because it is
possible to have repeated even degree factors. The vector space in
each degree is still finite dimensional.

The cohomology ring of the configuration space of based loops is just
the direct sum, $H^*(G^M;\R)=\bigoplus_{\pi_0(G^M)}
H^*((G^M)_0;\R)$. Notice that it is infinitely generated as an
algebra. The cohomology of the identity component will usually be the
important thing. Using the reductions, we see that the identity
component of the space of free maps is up to homotopy just the
product, $\free{M}{G}_0=G^M\times \hbox{Sp}(2)$, so the cohomology
ring $H^*(\free{M}{G}_0;\R)$ is obtained from $H^*((G^M)_0;\R)$ by
adjoining new generators in degrees $3$ and $7$, say $y_3$ and
$y_7$. Thus $H^2((G^M)_0;\Z)\cong H^2(\free{M}{G}_0;\Z)\cong
\Z\oplus\Z_2^3$.

\medskip

\noindent
{\bf Example 2} For this example, we take $M=T^3\# L(m,1)$,
$G_1=\hbox{SO} (8)$ and $G= \hbox{U}(2)\times \hbox{SO}(8)$. Recall
that the lens space $L(m,1)$ is the quotient $\hbox{Sp}(1)/\Z_m$ where
we view $\Z_m$ as the $m$-th roots of unity in
$S^1\subset\hbox{Sp}(1)$. In this example we will need to use
Reductions \ref{r6} and \ref{r7}. The unitary group is isomorphic to
$\hbox{Sp}(1)\times_{\Z_2} S^1$, where $\Z_2$ is viewed as the
diagonal subgroup, $\pm(1,1)$. The universal covering group of SO$(8)$
is Spin$(8)$. It follows that $G_1$ and $G$ are connected, $G$ has
universal covering group $\hbox{Sp}(1)\times\R\times\hbox{Spin}(8)$,
and fundamental groups are $\pi_1(\hbox{Spin}(8))=\Z_2$ and
$\pi_1(G)=\Z\oplus\Z_2$. The group $G$ has two simple factors, one of
which is symplectic. The integral cohomology of $M$ is given by
$H^1(M;\Z)\cong \Z^3$ and $H^2(M;\Z)\cong \Z^3\oplus\Z_m$. The
universal coefficient theorem and Proposition \ref{components} imply
$\pi_0(G_1^M)=\pi_0(\free{M}{G_1})=\Z\times\Z_2^4$ if $m$ is even,
$\Z\times\Z_2^3$ if $m$ is odd, and
$\pi_0(G^M)=\pi_0(\free{M}{G})=\Z^5\times\Z_2^4$ if $m$ is even and
$\Z^5\times\Z_2^3$ if $m$ is odd.  Theorem \ref{thm1} implies that
$\pi_1(G_1^M)=\Z^3\oplus\Z_m$,
$\pi_1(\free{M}{G_1})=\Z^3\oplus\Z_m\oplus\Z_2$,
$\pi_1(G^M)=\Z^6\oplus\Z_m^2\oplus\Z_2$ and
$\pi_1(\free{M}{G})=\Z^7\oplus\Z_m^2\oplus\Z_2^2$.

Turning once again to cohomology, we see from Theorem \ref{thm1co}
that  $H^*((\hbox{U}(2)^M)_0;\R)$ is the FUGCA generated by
$\mu([T^2\times \hbox{pt}]\otimes x_3)^{(1)}$, $\mu([S^1\times
\hbox{pt}\times S^1]\otimes x_3)^{(1)}$, $\mu([ \hbox{pt}\times
T^2]\otimes x_3)^{(1)}$, $\mu([S^1\times \hbox{pt}]\otimes
x_3)^{(2)}$, $\mu([ \hbox{pt}\times S^1 \hbox{pt}]\otimes x_3)^{(2)}$,
and $\mu([ \hbox{pt}\times S^1]\otimes x_3)^{(2)}$.

Also, $H^*((G_1^M)_0;\R)$ is the FUGCA generated by $\mu([T^2\times
\hbox{pt}]\otimes y_3)^{(1)}$, $\mu([S^1\times \hbox{pt}\times
S^1]\otimes y_3)^{(1)}$, $\mu([ \hbox{pt}\times T^2]\otimes
y_3)^{(1)}$, $\mu([S^1\times \hbox{pt}]\otimes y_3)^{(2)}$, $\mu([
\hbox{pt}\times S^1 \hbox{pt}]\otimes y_3)^{(2)}$, $\mu([
\hbox{pt}\times S^1]\otimes y_3)^{(2)}$, $\mu([T^3]\otimes
y_7)^{(4)}$, $\mu([T^3]\otimes z_7)^{(4)}$, $\mu([T^2\times
\hbox{pt}]\otimes y_7)^{(5)}$, $\mu([S^1\times \hbox{pt}\times
S^1]\otimes y_7)^{(5)}$, $\mu([ \hbox{pt}\times T^2]\otimes
y_7)^{(5)}$, $\mu([S^1\times \hbox{pt}]\otimes y_7)^{(6)}$, $\mu([
\hbox{pt}\times S^1 \hbox{pt}]\otimes y_7)^{(6)}$, $\mu([
\hbox{pt}\times S^1]\otimes y_7)^{(6)}$,  $\mu([T^2\times
\hbox{pt}]\otimes z_7)^{(5)}$, $\mu([S^1\times \hbox{pt}\times
S^1]\otimes z_7)^{(5)}$, $\mu([ \hbox{pt}\times T^2]\otimes
z_7)^{(5)}$, $\mu([S^1\times \hbox{pt}]\otimes z_7)^{(6)}$, $\mu([
\hbox{pt}\times S^1 \hbox{pt}]\otimes z_7)^{(6)}$, $\mu([
\hbox{pt}\times S^1]\otimes z_7)^{(6)}$, $\mu([T^3]\otimes
y_{11})^{(8)}$, $\mu([T^2\times \hbox{pt}]\otimes y_{11})^{(9)}$,
$\mu([S^1\times \hbox{pt}\times S^1]\otimes y_{11})^{(9)}$, $\mu([
\hbox{pt}\times T^2]\otimes y_{11})^{(9)}$, $\mu([S^1\times
\hbox{pt}]\otimes y_{11})^{(10)}$, $\mu([ \hbox{pt}\times S^1
\hbox{pt}]\otimes y_{11})^{(10)}$, and $\mu([ \hbox{pt}\times
S^1]\otimes y_{11})^{(10)}$.

Therefore, $H^*((G^M)_0;\R)$ is the FUGCA generated by all of the
generators listed for the two previous algebras. We changed the
notation for the generators of the cohomology of the lie groups as
needed. To get to the cohomology of the identity component of the
space of free maps, we would just have to add generators for the
cohomology of the group $G_0$ to this list. In general the cohomology
of a connected lie group is the same as the cohomology of the maximal
compact subgroup, and every compact Lie group has a finite cover that
is a product of simple, simply-connected, compact Lie groups and a
torus. In this case, we need to add generators, $t_1$, $u_3$, $w_3$,
$u_7$, $v_7$, and $u_{11}$. 

Thus, $H^2((G_1^M)_0;\Z)\cong\Z^6\oplus\Z_m$,
$H^2(\free{G_1}{M}_0;\Z)\cong\Z^6\oplus\Z_m\oplus\Z_2$,
$H^2((G^M)_0;\Z)\cong \Z^{21}\oplus\Z_m^2\oplus\Z_2$,   and
$H^2(\free{G}{M}_0;\Z)\cong \Z^{21}\oplus\Z_m^2\oplus\Z_2^2$.  We can
also analyze the topology of the space of $S^2$-valued maps with
domain $M$. The path components of $(S^2)^M$ agree with the path
components of $\free{M}{S^2}$ (Reduction \ref{r4}) and are given by Theorem
\ref{ponthm}. Let $\varphi_0:M\to S^2$ be the constant map and let
$\varphi_3:M\to S^2$ be the map constructed as the composition of the
map $M\to T^3$ (collapse the $L(m,1)$), the projection $T^3\to T^2$,
and a degree three map $T^2\to S^2$. According to Theorem \ref{thm2}
and Theorem \ref{freefun}, we have
$$
\begin{array}{rcl}
\pi_1(\free{M}{S^2}_{\varphi_0})=\pi_1((S^2)^M_{\varphi_0})&\cong&
\Z^6\oplus\Z_m\oplus\Z_2, \nonumber \\ 
\pi_1((S^2)^M_{\varphi_3})&\cong&
\Z^5\oplus\Z_m\oplus\Z_2, \qquad\hbox{and} \nonumber \\
\pi_1(\free{M}{S^2}_{\varphi_3})&\cong&
\Z^4\oplus\Z_m\oplus\Z_6\oplus\Z_2.
\end{array}
$$

Using Theorem \ref{thm2co} we can write out generators for the
cohomology. The cohomology, $H^*((S^2)^M_{\varphi_0};\R)$ is the FGCUA
generated by $PD([T^2\times \hbox{pt}])^{(1)} $, $PD([S^1\times
\hbox{pt}\times S^1]) ^{(1)} $, $PD([ \hbox{pt}\times T^2]) ^{(1)} $,
$\mu([T^2\times \hbox{pt}]\otimes x)^{(1)} $, $\mu([S^1\times
\hbox{pt}\times S^1]\otimes x)^{(1)}$, $\mu([ \hbox{pt}\times
T^2]\otimes x)^{(1)}$, $\mu([S^1\times \hbox{pt}]\otimes x)^{(2)}$,
$\mu([ \hbox{pt}\times S^1 \hbox{pt}]\otimes x)^{(2)}$, and $\mu([
\hbox{pt}\times S^1]\otimes x)^{(2)}$.

Similarly, $H^*((S^2)^M_{\varphi_3};\R)$ is the FGCUA generated by
$PD([S^1\times \hbox{pt}\times S^1]) ^{(1)} $, $PD([ \hbox{pt}\times
T^2]) ^{(1)} $,  $\mu([T^2\times \hbox{pt}]\otimes x)^{(1)} $,
$\mu([S^1\times \hbox{pt}\times S^1]\otimes x)^{(1)}$, $\mu([
\hbox{pt}\times T^2]\otimes x)^{(1)}$, $\mu([S^1\times
\hbox{pt}]\otimes x)^{(2)}$, $\mu([ \hbox{pt}\times S^1
\hbox{pt}]\otimes x)^{(2)}$, and $\mu([ \hbox{pt}\times S^1]\otimes
x)^{(2)}$.

The reason why there is no generator corresponding to $PD([T^2\times
\hbox{pt}])^{(1)}$ in the $\varphi_3$ cohomology is that it is not in
the kernel since $2\varphi_3^*\mu_{S^2}\cup PD([T^2\times
\hbox{pt}])^{(1)} = 6\mu_M$.

We can use Theorem \ref{freeco} to compute the cohomology of the space
of free maps. In the component with $\varphi_0$ we notice that the
second differential is trivial because $\varphi_0^*\mu_{S^2}=0$. It
follows that $H^*(\free{M}{S^2}_{\varphi_0};\R)$ is the
graded-commutative, unital algebra generated by    $PD([T^2\times
\hbox{pt}])^{(1)} $, $PD([S^1\times \hbox{pt}\times S^1]) ^{(1)} $,
$PD([ \hbox{pt}\times T^2]) ^{(1)} $,  $\mu([T^2\times
\hbox{pt}]\otimes x)^{(1)} $, $\mu([S^1\times \hbox{pt}\times
S^1]\otimes x)^{(1)}$, $\mu([ \hbox{pt}\times T^2]\otimes x)^{(1)}$,
$\mu([S^1\times \hbox{pt}]\otimes x)^{(2)}$, $\mu([ \hbox{pt}\times
S^1 \hbox{pt}]\otimes x)^{(2)}$, $\mu([ \hbox{pt}\times S^1]\otimes
x)^{(2)}$, and $\mu_{S^2}$. Notice that this algebra is not free. It
is subject to the single relation, $\mu_{S^2}^2=0$.

In the component containing $\varphi_3$ all of the generators of
$H^*((S^2)^M_{\varphi_3};\R)$ except $\mu([T^2\times \hbox{pt}]\otimes
x)^{(1)} $ survive to $H^*(\free{M}{S^2}_{\varphi_3};\R)$ because they
are in the kernel of $d_2$. However, $d_2\mu([T^2\times
\hbox{pt}]\otimes x)^{(1)}=6\mu_{S^2}$ so $\mu_{S^2}$ does not survive
and $H^*(\free{M}{S^2}_{\varphi_3};\R)$ is the FUGCA
generated by  $PD([S^1\times \hbox{pt}\times S^1]) ^{(1)} $, $PD([
\hbox{pt}\times T^2]) ^{(1)} $,  $\mu([S^1\times \hbox{pt}\times
S^1]\otimes x)^{(1)}$, $\mu([ \hbox{pt}\times T^2]\otimes x)^{(1)}$,
$\mu([S^1\times \hbox{pt}]\otimes x)^{(2)}$, $\mu([ \hbox{pt}\times
S^1 \hbox{pt}]\otimes x)^{(2)}$, and $\mu([ \hbox{pt}\times
S^1]\otimes x)^{(2)}$.

Thus $H^2((S^2)^M_{\varphi_0};\Z)\cong \Z^{18}\oplus\Z_m\oplus\Z_2$,
$H^2(\free{M}{S^2}_{\varphi_0};\Z)\cong \Z^{19}\oplus\Z_m\oplus\Z_2$,
$H^2((S^2)^M_{\varphi_3};\Z)\cong \Z^{13}\oplus\Z_m\oplus\Z_2$, and
$H^2(\free{M}{S^2}_{\varphi_3};\Z)\cong \Z^{9}\oplus\Z_m\oplus\Z_2$.

\section{Geometric interpretations}
\label{geo}
\news

We will follow the folklore maxim: think with intersection theory and
prove with cohomology.  The combination of Poincar\'e duality and the
Pontrjagin-Thom construction gives a powerful tool for visualizing
results in algebraic topology. If $W$ is an $n$-dimensional homology
manifold, Poincar\'e duality is the isomorphism $H^k(W)\cong
H_{n-k}(W)$. It is tempting to think of the $k$-th cohomology as the
dual of the $k$-th homology. This is not far from the truth. The
universal coefficient theorem is the split exact sequence 
$$
0\to
\hbox{Ext}^1_\Z(H_{k-1}(W;\Z),A)\to H^k(W;A)\to
\hbox{Hom}_\Z(H_k(W;\Z),A)\to 0. 
$$
Putting this together, we see that
every degree $k$ cohomology class corresponds to a unique $(n-k)$-cycle
(codimension $k$ homology cycle), and the image of the cocycle applied
to a $k$-cycle is the weighted number of intersection points with the
corresponding $(n-k)$-cycle. For field coefficients this is the entire
story since there is no torsion and the $\hbox{Ext}$ group
vanishes. With other coefficients, this gives the correct answer up to
torsion. The Pontrjagin-Thom construction associates a framed
codimension $k$ submanifold of $W$ to any map $W\to S^k$. The
associated submanifold is just the inverse image of a regular
point. This is well defined up to framed cobordism. Going the other way, a
framed submanifold produces a map $W\to S^k$ defined via the
exponential map on fibers of a tubular neighborhood of the submanifold
and as the constant map outside of the neighborhood. We will take this
up in greater detail later in this section. Before addressing the
topology of our configuration spaces, we need to understand the
cohomology of Lie groups.

A number of different approaches may be utilized to compute the real
cohomology of a compact Lie group:  H-space methods, equivariant Morse
theory, the Leray-Serre spectral sequence, Hodge theory. The
cohomology is a free graded-commutative algebra over ${\mathbb
R}$. Recall that this means that
$xy=(-1)^{\hbox{deg}(x)\hbox{deg}(y)}yx$. For our purposes, the
spectral sequence and Hodge theory are the two most important. The
fibration  $\hbox{SU}(N)\hookrightarrow\hbox{SU}(N+1)\to S^{2N+1}$ may
be used to compute the cohomology of SU$(N)$, and we will use it and
other similar fibrations to compute  the cohomology of various
configuration spaces. According to Hodge theory,  the real cohomology
is isomorphic to the collection of harmonic forms.  Any compact Lie
group admits an Ad-invariant innerproduct on the Lie algebra obtained
by averaging any innerproduct over the group, or as the Killing form,
$\langle X, Y \rangle=-\hbox{Tr}(\hbox{ad}(X)\hbox{ad}(Y))$ in the
semisimple  case. Such an innerproduct induces a biinvariant metric on
the group. With  respect to this metric, the space of harmonic forms
is  isomorphic to the space of Ad-invariant forms on the Lie
algebra. Any  harmonic form induces a form on the Lie algebra by
restriction and any Ad  invariant form on the Lie algebra induces a
harmonic form via left  translation.

In the case of SU$(N)$, these forms may be described as products of
the  elements, $x_j=\hbox{Tr}((u^{-1}du)^j)$. In some applications it
might be  appropriate to include a normalizing constant so that the
integral of each of these forms on an associated primitive homology
class is $1$. 

\subsection{Components of $G^M$}

For simplicity, we will just consider geometric descriptions of
 $G$-valued maps for the compact, simple, simply-connected Lie groups.
 By applying the Pontrjagin-Thom construction, we will obtain a
 correspondence between homotopy classes of based maps $M\ra G$ and
 finite collections of signed points in $M$. This may be used to give
 a geometric interpretation  of Proposition \ref{components}. In
 physical terms, the signed points may be  thought of as particles and
 anti-particles in the theory. 

To use the Pontrjagin-Thom construction in this setting we need a
special  basis for $H_*(G;\R)$.  By the universal coefficient theorem,
there are $(2k+1)$-cycles  $\beta_{2k+1}$ in $H_{2k+1}(G;{\mathbb R})$
dual to $x_{2k+1}$. Assuming that the generators $x_{2k+1}$ are
suitably  normalized, we may assume that the $\beta_{2k+1}$ are
integral classes i.e.\ images of elements of the form
$\beta^\prime_{2k+1}$ for  $\beta^\prime_{2k+1}\in H_{2k+1}(G;\Z)$. We
will often use notation from  de~Rham theory to denote the analogous
constructions in singular, or  cellular theory. For example, the
evaluation pairing between cohomology and  homology is called the cap
product. It is usually denoted, $x\cap\beta$ or  $x[\beta]$. The cap
product corresponds to integration ($\int_\beta x$) in  de Rham
theory. By Poincar\'e duality, we can identify each cocycle
$x_{2k+1}$ with a  codimension $(2k+1)$-cycle $F$ in $G$ so that the
image of any  $(2k+1)$-chain $c_{2k+1}$ under $x_{2k+1}$ is precisely
the algebraic intersection number of $F$ and $c_{2k+1}$. Hence, each
compact, simple, simply-connected Lie group contains a codimension $3$
cycle $F$ Poincar\'e dual to $x_3$, which intersects $\beta_{3}$
algebraically in one  positively oriented point.  We will shortly
describe these codimension 3 cycles in greater detail, but  we first
describe how these cycles may be used to determine the path
components of the configuration space. 

Assume for now that the cycle $F$ has a trivial normal bundle. We will
justify this assumption later.  (Throughout this paper we will use
normal bundles, open and closed tubular  neighborhoods  and the
relation between them via the exponential map  without explicitly
writing the map. If $\Sigma\subset M$ then  $\nu\Sigma\subset TM$,
will denote the normal bundle and  $N\Sigma\subset M$ will denote the
closed tubular neighborhood.)  Fix a trivialization of the normal
bundle. Using this trivialization, we  may associate a finite
collection of signed points to any generic based map,  $u:M\to G$.  To
such a map we associate the collection of points, $u^{-1}(F)$. Such a
point is positively oriented if the  push forward of an oriented frame
at the point has the same orientation as  the trivialization of the
normal bundle at the image. Conversely, to any  finite collection of
signed points we may associate a based map, $u:M\to G$.  Using a
positively or negatively oriented frame at each point, we construct  a
diffeomorphism from the closed tubular neighborhood of each point  to
the $3$-disk of radius $\pi$ in the space of purely imaginary
quaternions, ${\mathfrak sp}(1)$. Via the exponential map,
$\hbox{exp}:{\mathfrak sp}(1)\to \hbox{Sp}(1)$ given by,
exp$(x)=\cos(|x|)+\frac{\sin(|x|)}{|x|}x$ we define a map from the
closed tubular neighborhood of the points to Sp$(1)$. This map may be
extended to the whole $3$-manifold by sending points in the complement
of  the neighborhood to $-1$. We next modify the map by multiplying by
$-1$, so  that the base point will be $1$. Finally, we notice that the
class,  $\beta_3$ is represented by a homomorphic image of  $\Sp(1)$
in any Lie group. For the classical groups, this homomorphism is  just
the standard inclusion,  $\Sp(1)=\SU(2)\hookrightarrow\SU(n+1)$,
$\Sp(1)=\Spin(3)\hookrightarrow\Spin(n)$, or
$\Sp(1)\hookrightarrow\Sp(n)$. The homomorphism for each exceptional
group is described in \cite{AK1}. This matches exactly with the
statement  of Proposition \ref{components}. In the case we are
considering here,  $H_1(G;\Z)=0$, and an element of
$H^3(M;\pi_3(G))\cong H^3(M;H_3(\tilde G;\Z))\cong
H^3(M;H^{g-3}(\tilde G;\Z))$ is just a  machine that eats a $3$-cycle
in $M$, i.e. $[M]$, and spits out a machine  that eats a codimension
$3$-cycle in $G$, i.e. $F$, and spits out an integer.  If $G$ is not
simple, there will be independent codimension $3$-cycles for  each
simple factor, and one could interpret the intersection number with
each cycle as a different type of particle (soliton). If $G$ were not
simply connected, the element of $H^1(M;H_1(G_0))$ would be the
obvious one,  and one obtains the element of $H^3(M;\pi_3(G))$ from a
modification of the  map into $G$ that lifts to $\tilde G$.

It is not difficult to describe the cycles, $\beta_{2k+1}$ and $F$ for
$\SU(n+1)$. Recall that the suspension of a pointed topological space
is  $SX=X\times [0,1]/(X\times\{0,1\}\cup \{p_0\}\times [0,1])$. This
may be visualized as the product $X\times S^1$ with the circle  above
the marked point in $X$ and the copy of $X$ above a marked point in
$S^1$ collapsed to a point. Identify ${\mathbb C}P^k$ with
U$(k+1)/(\hbox{U}(1)\times\hbox{U}(k))$ and define $\beta_{2k+1}
:S{\mathbb C}P^k\to \hbox{SU}(n+1)$ by 
$$
\beta_{2k+1}([A,t])= [A, e^{\pi it}\oplus e^{-\pi it}I_k]_{\rm
com}\oplus I_{n-k}.
$$
Here, $[A,B]_{\rm com}=ABA^{-1}B^{-1}$ is the usual commutator in a
group.

The normalization constants of $x_{2k+1}$ would ensure that
$\int_{\beta_{2k+1}} x_{2k+1} =1$. The values of these constants for
$k=1$ have been computed in \cite{AK1}. We do not need these constants
for this  present work. The value of the normalization constants for
$k=2$ would, for  example, be  important if one wished to add a
Wess-Zumino term to the Skyrme Lagrangian.

The multiplication on a Lie group may be used to endow the homology of
the  Lie group with a unital, graded-commutative algebra structure,
and the  cohomology with a comultiplication. The homology product is
given by  $(\sigma:\Sigma\to G)\cdot(\sigma^\prime:\Sigma^\prime\to
G):=  (\sigma\sigma^\prime:\Sigma\times\Sigma^\prime\to G)$ and the
comultipication on cohomology is dual to this.  The multiplication and
comultiplication give  $H^*(G;{\mathbb R})$ the structure of a Hopf
algebra. It is exactly in this context that Hopf algebras were first
defined. Using this algebra structure, we may give an explicit
description of the Poincar\'e duality  isomorphism. Any product of
generators, $x_j$ in $H^*(G;{\mathbb R})$  is  sent to the element of
$H_*(G;{\mathbb R})$ obtained from the product,  $\prod_{k=1}^{n}
\beta_{2k+1}$ by removing the corresponding $\beta_j$. In  particular,
$F=\prod_{k=2}^{n} \beta_{2k+1}$ is the cycle  Poincar\'e dual to
$x_3$. Geometrically, Poincar\'e duality is described by the equation,
$\int_\Sigma \omega = \#(PD(\omega)\cap\Sigma)$.  Since $S{\mathbb
C}P^k$ is not a manifold some words about our  interpretation of the
normal bundle to $F$ are in order at this point. For  $\SU(2)$ we may
take $F=\{-1\}$. This is a codimension $3$ submanifold, so there are
no  problems. Recall that ${\mathbb C}P^k-{\mathbb C}P^{k-1}$ is
homeomorphic to  ${\mathbb R}^{2k}$. It follows that the subset of
$F$, call it $F_0$,  obtained from the product of the $S{\mathbb
C}P^k-S{\mathbb C}P^{k-1}$ is a  codimension $3$ cell  properly
embedded in $\SU(n+1)-(F-F_0)$. Since $F-F_0$ has codimension $5$,  we
may assume, using general position, that any map of a $3$-manifold
into  SU$(n+1)$ avoids $F-F_0$. As $F_0$ is contractible, it has a
trivial normal  bundle, justifying our assumption at the beginning of
this description.

\subsection{The fundamental group of $G^M$}\label{fundesc}

The Pontrjagin-Thom construction may also be used to understand the
isomorphism,
$$
\phi:\pi_1(G^M)\, \to\,{\mathbb Z}_2^s\,\oplus H^2(M; \pi_3(G)),
$$
asserted in Theorem \ref{thm1}. A loop in $(G^M)_0$ based at the
constant map $u(x)=1$, may be  regarded as a based map $\gamma:SM\to
G$. The identifications in the  suspension provide a particularly nice
way to summarize all of the constraints on $\gamma$ imposed by the
base points. We will use the same notation for the  map,
$\gamma:M\times [0,1]\to G$ obtained from $\gamma$ by  composition
with the natural projection.  The inverse image  $\gamma^{-1}(F)$ with
framing obtained by pulling  back the trivialization of $\nu(F)$ may
be associated to $\gamma$. Conversely,  given a framed link in
$(M-p_0)\times (0,1)$ one may construct an element  of
$\pi_1(G^M)$. Using the framing each fiber of the  closed tubular
neighborhood to the link may be identified with the disk of radius
$\pi$ in ${\mathfrak sp}(1)$. As before $-1$ times the  exponential
map may be used to construct  a map, $\gamma:SM\to G$ representing an
element of  $\pi_1(G^M)$.

It is now possible to describe the geometric content of the
isomorphism in  Theorem \ref{thm1}. For a class of loops $[\gamma]\in
(G^M)_0$, let $\phi(\gamma)=(\phi_1(\gamma),\phi_2(\gamma))$. Restrict
attention to the case of simply-connected $G$, and make the
identifications, $\pi_3(G)\cong  H_3(G;{\mathbb Z})\cong
H^{g-3}(G;{\mathbb Z})$. An element of  $H^2(M; \pi_3(G))$ may be
interpreted as a function that associates an integer to a surface in
$M$, say $\Sigma$, and a codimension $3$ cycle in $G$, say  $F$. Set
$\phi_2(\gamma)(\Sigma,F)=\#(\Sigma\times [0,1]\cap\gamma^{-1}(F))$.
Note that $\gamma^{-1}(F)$ inherits an orientation from the framing
and  orientation on $M$. Using Poincar\'e duality this may be said in
a  different way. The homology class of $\gamma^{-1}(F)$ in
$(M-p_0)\times (0,1)$ projects to an element of $H_1(M)$ dual to the
element associated to $\phi_2(\gamma)$. The first component of the
isomorphism counts the parity of the number of twists in the framing. 

Consider the framing in greater detail. Using a spin structure on $M$
we  associate a canonical framing to any oriented $1$-dimensional
submanifold of  $(M-p_0)\times (0,1)$. See Proposition \ref{split} in
the proofs  section. For now restrict attention to null-homologous
submanifolds. Let $\Sigma$ be an oriented $2$-dimensional
submanifold of $(M-N(p_0))\times  (0,1)$ with non trivial
boundary. The normal bundle to $\Sigma$ inherits an orientation from
the orientations on $(M-p_0)\times (0,1)$ and $\Sigma$.  Oriented
$2$-plane bundles are classified by the second cohomology. Since
$H^2(\Sigma;{\mathbb Z})=0$, the normal bundle is trivial. Let $(e_1,
e_2)$  be an oriented trivialization of this bundle. Let
$e_3\in\Gamma(T\Sigma|_{\partial\Sigma})$ be the outward unit
normal. The canonical framing on $\partial\Sigma$ is $(e_1, e_2,
e_3)$. Given  a second framing, $(f_1, f_2, f_3)$ on $\partial\Sigma$
and an orientation  preserving parameterization of the boundary, we
obtain an element  $A\in\pi_1(GL^+(3,{\mathbb
R}))=\pi_1(\hbox{SO}(3))\cong {\mathbb Z}_2$  satisfying $(f_1, f_2,
f_3)=(e_1, e_2, e_3)A$. This is the origin of the first component of
the isomorphism. The generator of  $\pi_1(\hbox{Sp}(1)^{S^3})\cong
{\mathbb Z}_2$ is represented by,
$$
\gamma:(\lambda,x_1,x_2)\mapsto\left(\begin{array}{cc}
x_1&-\bar{\lambda}\bar{x}_2 \\ \lambda x_2&\bar{x}_1\end{array}\right),
$$
having identified $S^3$ with the unit sphere in  ${\mathbb C}^2$ (so
$|x_1^2+|x_2|^2=1$), $S^1\cong$U$(1)$ and Sp$(1)\cong$SU$(2)$. The
image of $\gamma$ under the obvious inclusion
$\iota:$SU$(2)\rightarrow$SU$(3)$, that is,
$\iota(U)=\hbox{diag}(U,1)$,  is homotopically trivial, as can be seen
by constructing an explicit homotopy  between it and $\iota
\circ\gamma(1,\cdot)$. First note that any SU$(3)$  matrix is uniquely
determined by its first two columns, which must be an orthonormal
pair. For all $t\in [0,1]$, let  $\mu_t(\lambda)=t\lambda+1-t$ (so
$\mu_1=\hbox{id}$ and $\mu_0=1$) and define
$$
e:=\left(\begin{array}{c}x_1\\ \mu_t(\lambda)x_2\\
\sqrt{1-|\mu_t(\lambda)|^2}x_2\end{array}\right),\qquad
v:=\left(\begin{array}{c}-\bar{\lambda}\bar{x}_2\\ \bar{x}_1 \\ 0 
\end{array}\right),\qquad
v_\perp:=v-(e^\dagger v)e.
$$
Then
$$
(t,\lambda,x_1,x_2)\mapsto \left(e,\frac{v_\perp}{|v_\perp|},*\right)
$$
is the required homotopy between $\iota\circ\gamma$ ($t=1$) and the
trivial  loop based at $\iota:S^3\rightarrow $SU$(3)$. It is
straightforward to check  that $e$ and $v$ are never parallel (so the
map is well defined), that  $(t,\lambda,1,0)\mapsto I_3$ for all
$t,\lambda$ (this is a homotopy through  loops of based maps
$S^3\rightarrow$SU$(3)$) and that
$(t,1,x_1,x_2)\mapsto\iota(x_1,x_2)$ for all $t$ (each loop is based
at $\iota$). 

The homomorphic image of Sp$(1)$ is contained in a standardly embedded
SU$(3)$ in each of the exceptional groups and the classical groups
SU$(n+1)$,  $n\ge 2$, and Spin$(N)$, $N\ge 7$, \cite{AK1}. This is the
reason why the  ${\mathbb Z}_2$ factors only correspond to the
symplectic factors of the Lie group.

\begin{figure}
\hskip105bp\epsfig{file=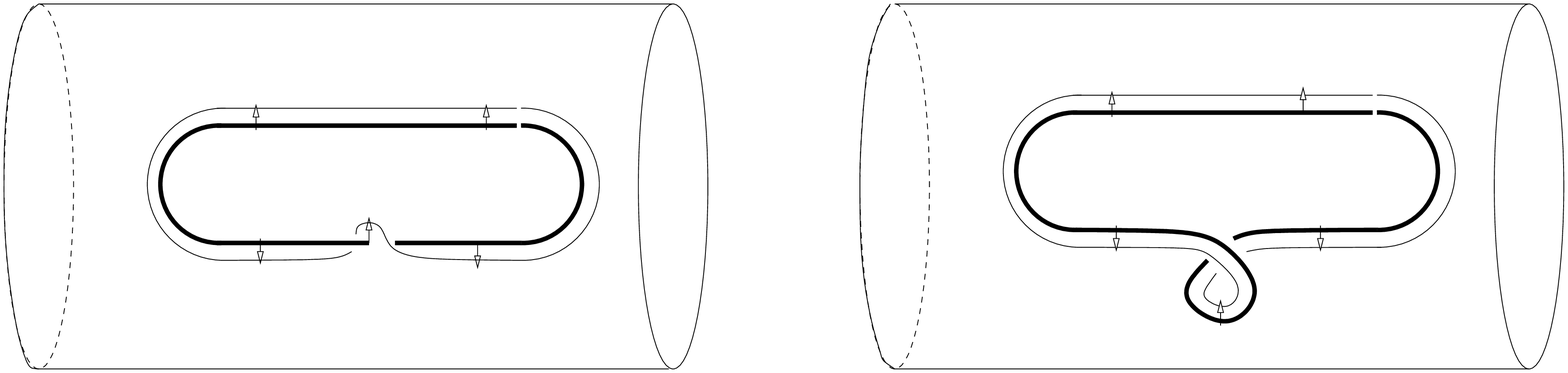,width=4truein}%
\caption{The rotation or exchange loop}\label{fig1}
\end{figure}

The following figures show some loops in the configuration spaces. For
the  first two figures, the horizontal direction represents the
interval direction  in $M\times [0,1]$. The disks represent the $x-y$
plane in a  coordinate chart in $M$, and we suppress the $z$ direction
due to lack of  space. Figure \ref{fig1} shows two copies of a typical
loop representing an  element in a symplectic ${\mathbb Z}_2$
factor. Only the first vector  of the framing is shown in figure
\ref{fig1}. The second vector is obtained by taking the cross product
with the tangent vector to the curve in the displayed slice, and the
final vector is the $z$-direction.  It is easy to see that the  left
copy may be deformed into the right copy. We describe the left copy as
follows: a particle and antiparticle are born; the particle undergoes
a full rotation; the two particles then annihilate.  The right copy
may be described  as follows: a first particle-antiparticle pair is
born; a second pair is born; the two particles exchange positions
without rotating; the first particle and  second antiparticle
annihilate; the remaining pair annihilates. Notice that  there are two
ways a pair of particles can exchange positions. Representing  the
particles by people in a room, the two people may step sideways
forwards/backwards and sideways following diametrically opposite
points on a  circle always facing the back of the room. This is the
exchange without  rotating described in figure \ref{fig1}. This
exchange is non-trivial in  $\pi_1(\hbox{Sp}(1)^{S^3})$. The second
way a pair of people may change  positions is to walk around a circle
at diametrically opposite points  always facing the direction that
they walk to end up facing the opposite  direction that they
started. This second change of position is actually  homotopically
trivial. Since the framed links in figure \ref{fig1}  avoid the
slices, $M\times \{0, 1\}$, they represent a loop based at the
constant identity map.

\begin{figure}
\hskip170bp\epsfig{file=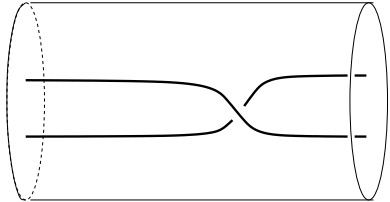,width=2truein}%
\caption{The degree $2$ exchange loop}
\label{fig2}
\end{figure}

It is possible to describe a framing without drawing any normal
vectors at  all. The first vector may be taken perpendicular to the
plane of the figure,  the second vector may be obtained from the cross
product with the tangent  vector, and the third vector may be taken to
be the suppressed $z$-direction.  The framing obtained by following
this convention is called the black board  framing. We use the
blackboard framing in figure \ref{fig2}. The  Pontrjagin-Thom
construction may also be used to visualize loops in other  components
of the configuration space. Figure \ref{fig2} shows a loop in the
degree $2$ component of the space of maps from $M$ to Sp$(1)$.

\begin{figure}
\hskip170bp\epsfig{file=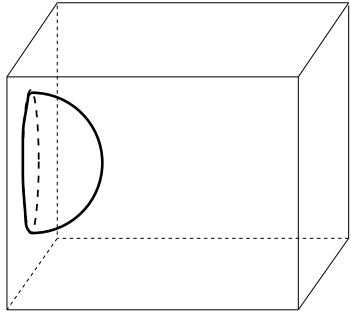,width=1.8truein}%
\caption{The contraction of a canonically framed contractible link}
\label{fig3}
\end{figure}

We can also use the Pontrjagin-Thom construction to draw figures of
homotopies between loops in configuration space. Figure \ref{fig3}
displays a homotopy  between the loop corresponding to a canonically
framed unknot and the constant loop. In this figure, the horizontal
direction represents the second interval  factor of $M\times
[0,1]\times [0,1]$, the direction out of the page  represents the
first interval factor, the vertical direction represents the  $x$
direction, and the $y$ and $z$ directions are suppressed. The framing
is  given by the normal vector to the hemisphere, the $y$ direction
and the $z$  direction.

\subsection{Cohomology of $G^M$}\label{cohomdesc}

We now turn to a description of the real cohomology of $G^M$. We will
use the slant product to associate a cohomology class on $G^M$ to a
pair consisiting of a homology class on $M$ and a cohomology class on
$G$.  Recall that the slant product is a map $H^n(X\times Y;A)\otimes
H_k(X;B)\to H^{n-k}(Y;A\otimes B)$, \cite{span}. In addition the
universal coefficient theorem allows us to identify $H^k(G^M;\R)$ with
$\hbox{Hom}(H_k(G^M;\Z),\R)$.  Let $\sigma:\Sigma\to M$ be a singular
chain representing a homology class in $H_d(M;\R)=H_d(M;\Z)\otimes \R$
(instead of viewing singular chains as linear combinations of singular
simplicies, we will combine them together and view a singular chain as
a map of a special polytope into the space), and let $x_j$ be a
cohomology class in $H^j(G;\R)$. To define the image of the mu map,
$\mu(\Sigma\otimes x_j)$, let $u:F\to G^M$ be a singular chain
representing an element in the $H_{d-j}(G^M)$. This induces a natural
singular chain $\widehat u:M\times F\to G$.  The pull back produces
$\widehat u^*x_j\in H^j(M\times F;\R)$. The formal definition of the
mu map is then, 
\beq
\label{mudef} 
\mu(\Sigma\otimes x_j)(u):=(\widehat
u^*x_j/\Sigma)[F].  
\eeq 
Writing this in notation from the de Rham
model of cohomology may help to clarify the definitions. In principle
one could construct a homology theory based on smooth chains and make
the following rigorous. The $\mu$ map produces a $(j-d)$-cocycle in
$G^M$ from a $d$-cycle in $M$ and a $j$-cocycle in $G$.  On the level
of chains, let $e^d:D^d\to M$ be a $d$-cell, and $x_j$ be a  closed
$j$-form on $G$. Given a singular simplex, $u:\Delta^{j-d}\to G^M$,
let $\widehat u:M\times \Delta^{j-d}\to G$ be the natural map and write
$$
\mu(e^d\otimes x_j)(u)= \int_{D^d\times \Delta^{j-d}} \widehat u^* x_j.
$$
Using the product formula for the boundary,
$$
\partial(D^d\times\Delta^{j-d+1}) = (\partial
D^d)\times\Delta^{j-d+1} + (-1)^{d}D^d\times \partial \Delta^{j-d+1},
$$
we can get a simple formula for  the coboundary of the image of an
element under the $\mu$-map. Let  $v:\Delta^{j-d+1}\to G^M$, be a
singular simplex, then 
\bea 
\delta(\mu(e^d\otimes
x_j))(v)&=&\sum_{k=0}^{j-d+1} (-1)^k \int_{D^d\times  \Delta^{j-d}}
\widehat{(v\circ f_k)}^* x_j = \int_{D^d\times
\partial\Delta^{j-d+1}} \widehat v^* x_j \nonumber \\ &=&
(-1)^{d+1}\int_{(\partial D^d)\times \Delta^{j-d+1}} \widehat v^*
x_j+(-1)^d\int_{\partial(D^d\times\Delta^{j-d+1})}\widehat v^*
x_j\nonumber\\
\label{delta}
&=& (-1)^{d+1}\mu((\partial e^d)\otimes x_j)(v).  
\eea 
We used Stokes'
theorem in the last line. It follows that $\mu$ is well  defined at
the level of homology. Theorem \ref{thm1co} asserts that
$H^*(G^M_0;\R)$ is a finitely generated algebra with generators
$\mu(\Sigma_j^d \otimes x_k)$ where $\{\Sigma_j^d\}$ and $\{x_k\}$ are
bases for $H_*(M;\R)$ and $H^*(G;\R)$ respectively. The multiplication
on $H^*(G^M;\R)$ is given by the cup product. Recall that this is
defined at the level of  cochains by,
$(\alpha\cp\beta)(w)=\alpha(~_kw) \beta(w_\ell)$, where $\alpha$ is a
$k$-cocycle, $\beta$ is a  $\ell$-cocycle, $w$ is a
$(k+\ell)$-singular simplex and $~_kw$ is the front $k$-face and
$w_\ell$ is the back $\ell$-face \cite{span}. Note that $\cp$ is
graded-commutative, that is, $\alpha\cp\beta
=(-1)^{k\ell}\beta\cp\alpha$. 

It is instructive to understand some classes that do not appear as
generators. One might expect $\mu(\hbox{pt}\otimes x_j)$ to be a
generator in degree $j$. However, since $G^M$  consists of based maps,
the induced  map $\widehat u:M\times F\to G$ arising from a chain
$u:F\to G^M$ restricts to a constant map on $\hbox{pt}\times F$. It
follows that $\mu(\hbox{pt}\otimes x_j)=0$. There would be an
analogous class if we considered the cohomology of the space of free
maps. Turning to the other end of the spectrum, one might expect to
see classes of the form $\mu(M\otimes x_3)$ in degree zero.  Such
certainly could not appear in the cohomology of the identity component
$G^M_0$. In fact we stated our theorem for the identity component
because the argument leading to generators of the form
$\mu(\Sigma\otimes x_3)$ breaks down when $\Sigma$ is a $3$-cycle and
$x_3$ is a $3$-cocycle. The argument starts by considering maps of
spheres into the group $G$, and then assembles the cohomology of these
mapping spaces (which are denoted by $\Omega^kG$) into the cohomology
of $G^M$. The path fibration is used to compute the cohomology of the
$\Omega^kG$. The fibration leading to the cohomology of $\Omega^3G$
does not have a simply connected base and this is the break down. See
Lemma \ref{og}. Finally one might expect to see classes of the form
$\mu(\Sigma\otimes x_j\cup x_k)$. It will turn out in the course of
the proof (Lemma \ref{og}) that such classes vanish.

Up to this point, our geometric descriptions of the algebraic topology
of configuration spaces have  been simpler than we had any right to
expect. We were able to describe the  space of path components and the
fundamental group of the  configuration space of maps from an
orientable $3$-manifold into an arbitrary  simply-connected Lie group
by just considering subgroups isomorphic to  Sp$(1)$. This will not
hold for all homotopy invariants  of $G^M$. The main object of
interest to us is the second  cohomology of the configuration space
with integral coefficients, because  this classifies the complex line
bundles over the configuration space (the quantization
ambiguitity). It is  possible to describe one second cohomology class
on Sp$(n)^M$ in terms of  Sp$(1)$ geometry. However we need to pass to
SU$(3)$ subgroups to get at the  second cohomology in general. 

Before considering these geometric representatives of the second
cohomology, briefly recall the definition of the Ext groups. Given
$R$-modules $A$ and $B$  pick a free resolution of $A$ say $\to C_2\to
C_1\to C_0\to A$. The $k$th Ext group is just defined to be the $k$th
homology of the complex $\hbox{Hom}(C_*,B)$,
i.e. $\hbox{Ext}^k_R(A,B)=H_k(\hbox{Hom}(C_*,B))$.  When $R$ is a PID
(principal ideal domain)
every $R$-module has a free resolution of the form, $0\to C_1\to
C_0\to A$. Given such a resolution one obtains the exact sequence,
\beq
\label{ext} 0\to \hbox{Hom}(A,B)\to \hbox{Hom}(C_0,B)\to
\hbox{Hom}(C_1,B)\to \hbox{Ext}^1_R(A,B)\to 0, 
\eeq 
and all higher Ext
groups vanish. We will always take $R=\Z$ and drop the ground ring
from the notation. Based on the above exact sequence, we say that the
Ext groups measure the failure of Hom to be exact i.e. take exact
sequences to exact sequences. The Ext group may also be identified
with the collection of extensions of $A$ by $B$ \cite{span}.  By the
universal coefficient theorem  
$$
\begin{array}{rcl}
H^2(G^M;\Z)&\cong&\hbox{Ext}^1(H_1(G^M;\Z),\Z)\oplus
\hbox{Hom}(H_2(G^M;\Z),\Z)
 \\
H^2(G^M;\R)&\cong&\hbox{Ext}^1(H_1(G^M;\Z),\R)\oplus
\hbox{Hom}(H_2(G^M;\Z),\R)
\end{array}
$$ 
Now for all $A$, $\hbox{Ext}^1(A,\R)=0$, so
$\hbox{Hom}(H_2(G^M;\Z),\Z)$ is a free abelian group of rank
$b_2=\hbox{dim}_{\R}H^2(G^M;\R)$. In addition, $\hbox{Ext}^1(A,\Z)$ is
just  the torsion subgroup of $A$ and $H_1(G^M;\Z)\cong
\pi_1^{\rm ab}(G^M)= \pi_1(G^M)$. Hence 
$$
H^2(G^M;\Z)\cong\Z^{b_2}\oplus\tor(\pi_1(G^M)) 
$$ where $\pi_1(G^M)$ and
the Betti number $b_2$ may be obtained from Theorems \ref{thm1} and
\ref{thm1co}.

We will use the universal coefficient theorem and Ext groups to
describe some cohomology classes of our configuration spaces.
There is a natural ${\mathbb Z}_2$ contained in
the fundamental group of the  configuration space for any group with a
symplectic factor. This ${\mathbb Z}_2$ is  generated by the exchange
loop. Wrapping twice around the exchange  loop is the boundary of a
disk in the configuration space. Since  ${\mathbb R}P^2$ is the result
of identifying the points on the boundary of a  disk via a degree $2$
map, one expects to find an ${\mathbb R}P^2$ embedded  into any of the
Skyrme configuration spaces with a symplectic factor. In
\cite{sorkin}, R.~Sorkin  describes an embedding,
$f_{\rm{stat}}:{\mathbb R}P^2\to \hbox{SU}(n+1)^{S^3}$. He also
describes an embedding, $f_{\rm{spin}}:{\mathbb R}P^3\to
\hbox{SU}(n+1)^{S^3}$. He  further shows that $f_{\rm{spin}}$
restricted to the ${\mathbb R}P^2$  subspace is homotopic to
$f_{\rm{stat}}$. Using the map,  $M\to M^{(3)}/M^{(2)}\cong S^3$,
these induce maps into SU$(n+1)^M$. Here $M^{(k)}$ is the $k$ skeleton
of $M$ with respect to some CW structure. 
In fact, using the inclusion of Sp$(1)=\hbox{SU}(2)$ into any
simply-connected simple Lie group one obtains maps from $\RP^2$ and
$\RP^3$ into any configuration space of Lie group valued maps. This is
most interesting when the map factors through a symplectic factor.

We briefly recall Sorkin's elegant construction. Describe
${\mathbb R}P^2$ as the $2$-sphere with antipodal points
identified. By the addition of particle  antiparticle pairs, we may
assume that there are two particles  in a coordinate chart. We may
place the particles at antipodal points of a  sphere in a coordinate
chart using frames parallel to the coordinate  directions. The map
obtained from these frames using the Pontrjagin-Thom  construction is
$f_{\rm{stat}}$. The projective space, ${\mathbb R}P^3$ is
homeomorphic to the rotation group SO$(3)$. The map $f_{\rm{spin}}$
may be  described by using SO$(3)$ to rotate a single frame and then
applying the Pontrjagin-Thom construction. Sorkin includes a second
unaffected particle in his description of $f_{\rm{spin}}$ to make
the comparison with  $f_{\rm{stat}}$ easier.

A degree one map $M\to S^3$ (which always exists) induces a map
$G^{S^3}\to G^M$. The space $G^{S^3}$ is typically denoted
$\Omega^3G$. If the Lie algebra of the maximal compact subgroup admits
a symplectic factor, then we have an interesting map Sp$(1)\to G$
which induces a map $\Omega^3\hbox{Sp}(1)\to \Omega^3G$. We will see
in the course of our proofs that on the level of $\pi_1$ or $H_1$
these maps give a sequence of injections,
$$
H_1(\RP^2;\Z)\to H_1(\RP^3;\Z)\to H_1(\Omega^3\hbox{Sp}(1);\Z)\to
H_1(\Omega^3G;\Z)\to H_1(G^M;\Z).
$$
The universal coefficient theorem implies that there is a $\Z_2$
factor in the second cohomology of $G^M$ when $G$ contains a
symplectic factor. In fact, in this case we see that twice the
exchange loop is a generator of the $1$-dimensional boundaries. This
means we can define a homomorphism from $B_1$ (the $1$-dimensional
boundaries) to $\Z$ taking twice the exchange loop to $1$. The cocycle
defined by following the boundary map from $2$-chains by this
homomorphism generates the $\Z_2$ in $H^2(G^M;\Z)$. We see that this
class evaluates nontrivially on the Sorkin $\RP^2$.

When $G$ has unitary factors, there will be infinite cyclic factors in
the second cohomology of $G^M$. This is nicely explained by a
construction of Ramadas, \cite{ramadas}. Ramadas constructs a map,
$S^2\times S^3\to \hbox{SU}(3)$. This construction goes as follows. He
first defines a map $K: \hbox{SU}(2)\to \hbox{SU}(2)$ by
$$
K(\left(\begin{array}{cc}a&b\\-\bar b&\bar a\end{array}\right)
)=(|a|^4+|b|^4)^{-\frac12}\left(\begin{array}{cc}a^2&-\bar b^2\\
b^2&\bar a^2\end{array}\right).
$$
This map satisfies,
$K(\hbox{diag}(\lambda,\bar\lambda)A)=K(A)\hbox{diag}(\lambda^2,
\bar\lambda^2)$.
Finally define $\widehat\sigma: S^1\backslash \hbox{SU}(2)
\times\hbox{SU}(2)\to\hbox{SU}(3)$ by 
$$
\widehat\sigma([A],B)=
\hbox{diag}(1,K(A)) \hbox{diag}(ABA^*,1)\hbox{diag}(1,K(A)^*).
$$
Here we are
viewing $S^1$ as the subgroup $\hbox{diag}(\lambda,\bar\lambda)$ of
SU$(2)$. It is well known that $ S^1\backslash \hbox{SU}(2)\cong S^2$
and SU$(2)\cong S^3$. The map $\widehat\sigma:S^2\times S^3\to
\hbox{SU}(3)$ induces a map
$\sigma:S^2\to\Omega^3\hbox{SU}(3)$. Ramadas shows that this map
generates $H_2(\Omega^3\hbox{SU}(3);\Z)\cong \Z$. Combining with the
degree one map from $M$ and the inclusion into a special unitary
factor of $G$, we obtain a map $S^2\to G^M$ generating an
infinite cyclic factor of $H_2(G^M;\Z)$. By the universal coefficient theorem
a map from $H_2(G^M;\Z)$ to $\Z$ taking this generator to $1$ is a cohomology
class in $H^2(G^M;\Z)$. Clearly this class evaluates non-trivially on
this $S^2$.

If $G$ does not have a symplectic or special unitary factor, then
there is no reason to expect any elements of the second cohomology. In
fact under this hypothesis, $H^2(\Omega^3G;\Z)=0$. It is worth
mentioning how these maps behave in general. The third homotopy group
of any Lie group is generated by homomorphic images of Sp$(1)$. Each
time one of these generators is contained in a symplectic factor, we
get a $\Z_2$ in the second cohomology detected by a Sorkin
$\RP^2$. When one of these factors is not contained in a symplectic
factor, it is contained in a copy of SU$(3)$. This kills the $\Z_2$
factor in $\pi_1$ as explained above in subsection \ref{fundesc}. If
the SU$(3)$ is contained in a special unitary factor, the Sorkin map
$\RP^2\to \hbox{Sp}(1)^M\to \hbox{SU}(3)^M \to G^M$ pulls back the
second cohomology class described by Ramadas (and extended to
arbitrary $M$ and $G$ with special unitary factor as above) to the
generator of $H^2(\RP^2;\Z)\cong \Z_2$. (Ramadas proves that the
generator of $H^2(\Omega^3\hbox{SU}(3);\Z)$ pulls back to the
generator of $H^2(\RP^2;\Z)$ and the rest follows from our proofs.)
 If this
SU$(3)$ is not contained in a special unitary factor, it follows from
our proofs that the second homology class associated to $S^2\to G^M$
bounds, so there is no associated cohomology class.  

\subsection{Components of $(S^2)^M_\varphi$}\label{scomp}

\begin{figure}
\hskip95bp\epsfig{file=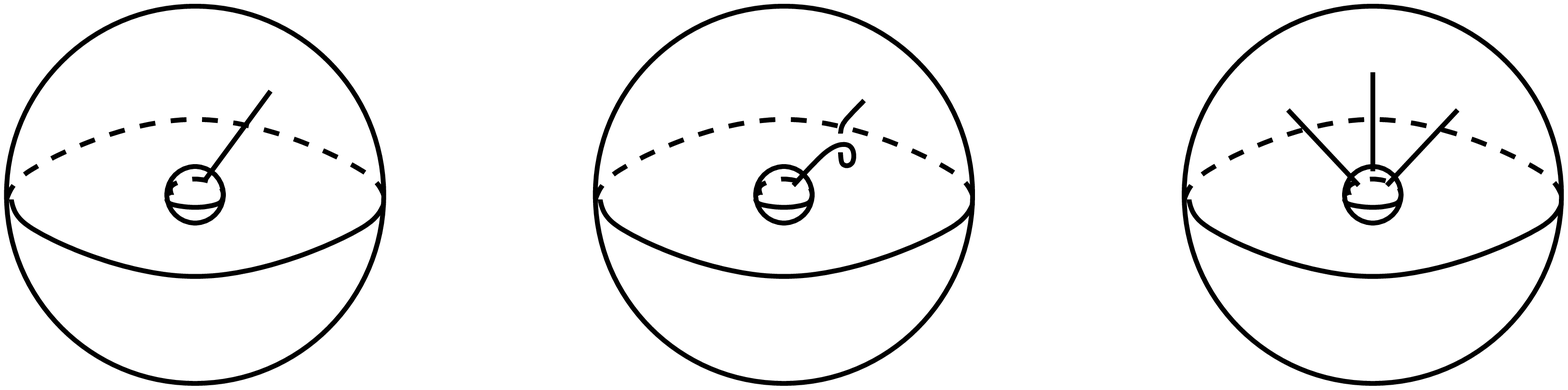,width=4truein}%
\caption{Pontrjagin-Thom representatives of $S^2$-valued
maps}\label{sputnik}
\end{figure}

\begin{figure}
\hskip80bp\epsfig{file=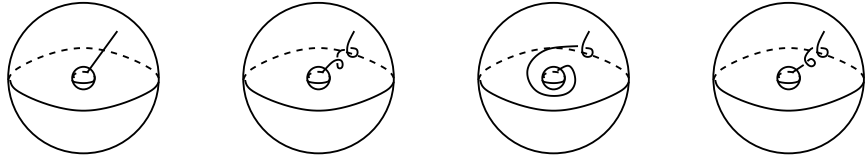,width=4.5truein}%
\caption{Introducing $2d$ twists}\label{untwist}
\end{figure}

The picture of the components of $(S^2)^M_\varphi$ arising from the
Pontrjagin-Thom construction and Poincar\'e duality is quite nice. The
inverse image of a regular point in $S^2$ is Poincar\'e dual to
$\varphi^*\mu_{S^2}$. The number of twists in the framing of a second
map with the same pull-back is the element of $H^3(M;\Z)/\langle
2\varphi^*\mu_{S^2}\rangle$. This is very similar to the description
of elements of the fundamental group of $G^M$ when $G$ has symplectic
factors. We will identify $S^2$ with $\CP^1$ and consider several
maps. We have $\varphi_1, \varphi_1^\prime, \varphi_3:\CP^1\times
S^1\to \CP^1$ given by,
$$
\varphi_1([z:w],\lambda)= [z:w],\qquad
\varphi_1^\prime([z:w],\lambda)=[\lambda z:w],\quad\hbox{and}\quad
\varphi_3([z:w],\lambda)=[z^3:w^3].
$$
We can view $\CP^1\times S^1$ as $S^2\times [0,1]$ (a spherical shell)
with the inner and outer ($S^2\times\{0\}$ and $S^2\times\{0\}$)
spheres identified. Using this convention and the framing conventions
from subsection \ref{fundesc},  We have displayed the framed
$1$-manifolds arising as the inverse image of a regular point in
figure \ref{sputnik}.
It may appear that there is a well defined twist number associated to
a $S^2$-valued map. However, there is a homeomorphism of $\CP^1\times
S^1$ twisting the $2$-sphere (such a map is given by
$([z:w],\lambda)\mapsto ([\lambda z:w],\lambda)$). This will change
the number of twists in a framing, but will not change the relative
number of twists. The reason why this relative number of twists is
only well defined modulo twice the divisibility  of the cohomology
class $\varphi^*\mu_{S^2}$ is demonstrated for $\varphi_1$ in figure
\ref{untwist}.

\subsection{Fundamental group of $(S^2)^M_\varphi$}

An element of $\pi_1((S^2)^M_\varphi)$ is represented by a map,
$\gamma:M\times S^1\to S^2$. The inverse image of a regular point is a
$2$-dimensional submanifold, say $\Sigma$. This defines an element of
$H^1(M;\Z)$ as follows. To any $1$-cycle in $M$, say $\sigma$, we
associate the intersection number of $\Sigma$ and $\sigma\times
S^1$. Since our loop is in the path component of $\varphi$, the
surface $\Sigma$ is parallel to the $\varphi$-inverse image of a
regular point. This implies that our element of $H^1(M;\Z)$ is in the
kernel of the map, $2\varphi^*\mu_{S^2}:H^1(M;\Z)\to H^3(M;\Z)$. Given
any element of this kernel, we can define a loop in $(S^2)^M_\varphi$
via the ${\mathfrak q}$-map defined below at line \ref{qdef}. There is
a map from $u:M\times S^1\to \hbox{Sp}(1)$ that may be used to change
this new loop back into $\gamma$. The remaining homotopy invariants of
$\gamma$ are just those of $u$ as described in subsection
\ref{fundesc}.

\section{Physical consequences}
\label{phy}
\news

As explained in section \ref{reduct}, the configuration space of the Skyrme 
model with arbitrary
target group is homotopy equivalent to the configuration space of a
collection of uncoupled Skyrme fields each taking values in a
compact, simply connected, simple Lie group.  We
will therefore assume, throughout this section that $G$ is compact, simply 
connected
and simple. In this case, by Proposition \ref{components}, the
path components of $G^M$ are labeled by $H^3(M;\Z)\cong\Z$, identified with
the baryon number $B$ of the configuration. This identification has already
been justified by consideration of the Pontrjagin-Thom construction.
Let us denote the baryon number $B$ sector by $Q_B$. 

We first recall how Finkelstein and Rubinstein introduced fermionicity to
the Skyrme model \cite{finrub}. 
The idea is that the quantum state is specified by
a wavefunction on $\wt{Q}_B$, the universal cover of $Q_B$, rather than
$Q_B$ itself. By the uniqueness of lifts, there is a natural action of
$\pi_1(Q_B)$ on $\wt{Q}_B$ by deck transformations. Let $\pi:\wt{Q}_B\ra
Q_B$ denote the covering projection, $\lambda\in \pi_1(Q_B)$ and $D_\lambda$
be the associated deck transformation. Since all points in $\pi^{-1}(u)$
are physically indistinguishable, we must impose the constraint
$$
|\psi(D_\lambda q)|=|\psi(q)|
$$
on the wavefunction $\psi:\wt{Q}_B\ra \C$, for all $q\in \wt{Q}_B$ and
$\lambda\in \pi_1(Q_B)$. This leaves us the freedom to assign phases
to the deck transformations, that is, the remaining quantization ambiguity
consists of a choice of $U(1)$ representation of $\pi_1(Q_B)$. The 
possibility of fermionic quantization arises if the two-Skyrmion exchange
loop in $Q_2$ is noncontractible with even order: we can then choose
a representation which assigns this loop the phase $-1$. In this case our 
wavefunction aquires a minus sign under Skyrmion interchange. Clearly, the Finkelstein-Rubinstein model could apply to any sigma model with a configuration space admiting non-trivial elements of the fundamental group representing the exchange of identical particles. In particular, the domain does not have to be $\R^3$. 

Note we have insisted that the wavefunction $\psi$ have support on a single 
path component
$Q_B$, because baryon number is conserved in nature, so transitions which 
change $B$ have zero probability. It seems, then, that the choice of
representation of $\pi_1(Q_B)$ can be made independently for each $B$, but
in fact there is a strong consistency requirement between the 
representations associated with the various components.
Recall that all the sectors are homeomorphic and
that given any $u\in Q_B$ one obtains a homeomorphism $Q_0\ra Q_B$ by
pointwise multiplication by $u$. Hence, to each $u\in Q_B$ there is 
associated an isomorphism $\pi_1(Q_0)\ra\pi_1(Q_B)$, so one has a map
$Q_B\ra {\rm Iso}(\pi_1(Q_0),\pi_1(Q_B))$. Since $Q_B$ is connected and 
$\pi_1$
is discrete, this map is constant, that is, there is a {\em canonical}
isomorphism $\pi_1(Q_0)\ra\pi_1(Q_B)$, which may be obtained by pointwise
multiplication by {\em any} charge $B$ configuration. Having chosen a
representation of $\pi_1(Q_0)$, we obtain canonical representations of
$\pi_1(Q_B)$ for all other $B$. Physically, we are demanding that the
phase introduced by transporting a configuration around a closed loop 
should be independent of the presence of static Skyrmions held remote
from the loop. This places nontrivial consistency conditions, if we are
to obtain a genuinely fermionic quantization. In particular, the loop in
$Q_{2B}$ consisting of the exchange of a pair of identical charge $B$
Skyrmions must be assigned the phase $(-1)^B$, since a charge $B$ Skyrmion
represents a bound state of $B$ nucleons, which is a fermion for $B$ odd
and a boson for $B$ even. 

\begin{figure}
\begin{center}
\epsfig{file=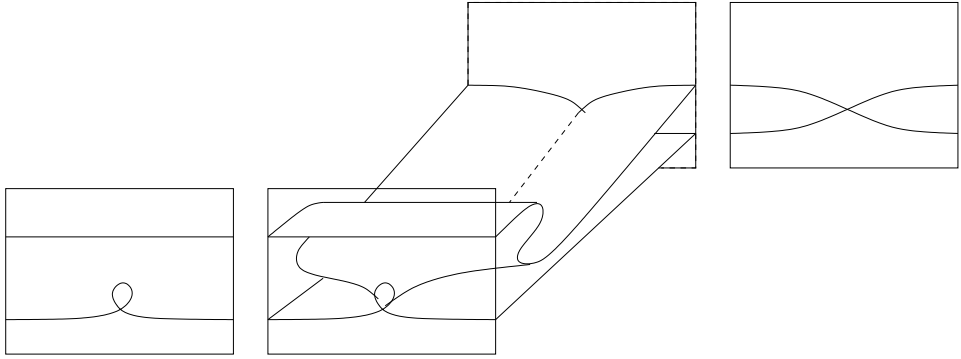,width=5truein}%
\caption{ The cobordism between Skyrmion rotation and Skyrmion
exchange.  }
\label{fighwb}
\end{center}
\end{figure}

The Finkelstein-Rubinstein formalism can be used to give a consistent
fermionic quantization of the Skyrme model on any domain $M$ if
$G=\Sp(n)$, but not for any of the other simple target groups. In this
case, Theorem \ref{thm1} tells us that 
$$ 
\pi_1(Q_B)\equiv\Z_2\oplus H_1(M) 
$$
and we can choose (and fix) a $U(1)$ representation which maps the
generator of $\Z_2$ to $(-1)$. The generator of the $\Z_2$-factor in the baryon number zero component is exactly the rotation--exchange loop as may be seen in the proof of proposition \ref{prop1} in the next section. To see that this assigns phase $(-1)$
to the 2-Skyrmion exchange loop, we may consider the Pontrjagin-Thom
representative of the loop. This is a framed 1-cycle in $S^1\times M$
depicted in figure \ref{fighwb}. It is framed-cobordant to the
representative of the loop in which one of the Skyrmions remains
static, while the other rotates through $2\pi$ about its center. 
Figure \ref{fighwb} gives a sketch of the cobordism. The
horizontal direction represents the loop parameter (``time''), the vertical
direction represents $M$ and the direction into the page represents the
cobordism parameter. The framing has been omitted, and the start and
end 1-cycles of the cobordism have been repeated, for clarity. Note that the
apparent self intersection of the cobordism (along the dashed line) is an
artifact of the pictorial projection from 5 dimensions to 3. Hence, the  
exchange loop in
$Q_2$ is homotopic to the loop represented by one static Skyrmion and one Skyrmion that undergoes a full rotation. 

\begin{figure}
\begin{center}
\epsfig{file=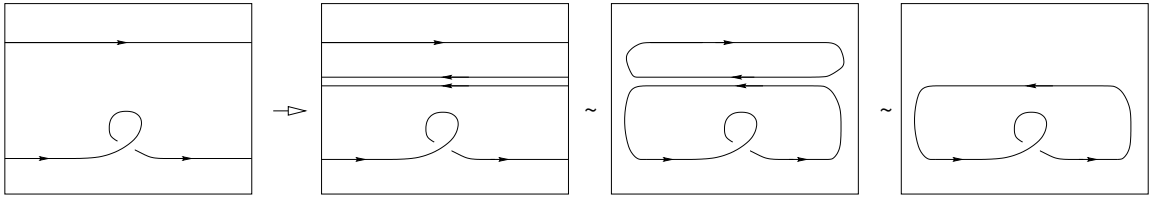,width=6truein}%
\caption{ Mapping the Skyrmion exchange loop into the vacuum sector
$Q_0$.  }
\label{figexch1}
\end{center}
\end{figure}

To identify the phase assigned to this homotopy
class, we must transfer the loops to $Q_0$ by adding a pair of
anti-Skyrmions, as depicted in figure \ref{figexch1}. This changes
each configuration by multiplying by a fixed charge $-2$ configuration
which is $1$ outside a small ball -- precisely one of the
homeomorphisms discussed above. 
The figure may be described thus: the
exchange loop is homotopic to the rotation loop with an extra static $1$-Skyrmion lump (far left) which is 
transferred to the vacuum sector by adding a stationary pair of 
anti-Skyrmions (2nd box). This loop is homotopic to the charge $0$  rotation loop of
figure \ref{fig1}, via the sequence of moves shown. The orientations on the
curves indicate how to assign a framing via the blackboard framing 
convention.
The resulting  Pontrjagin-Thom
representative is framed cobordant to the charge $0$ exchange loop
described in section \ref{fundesc}, which, as explained, generates the
$\Z_2$ factor in $\pi_1(Q_0)$.  Hence, the loop along which two
identical $1$-Skyrmions are exchanged (without rotating) around a contractible path
in $M$ is assigned the phase $(-1)$.  

\begin{figure}
\begin{center}
\epsfig{file=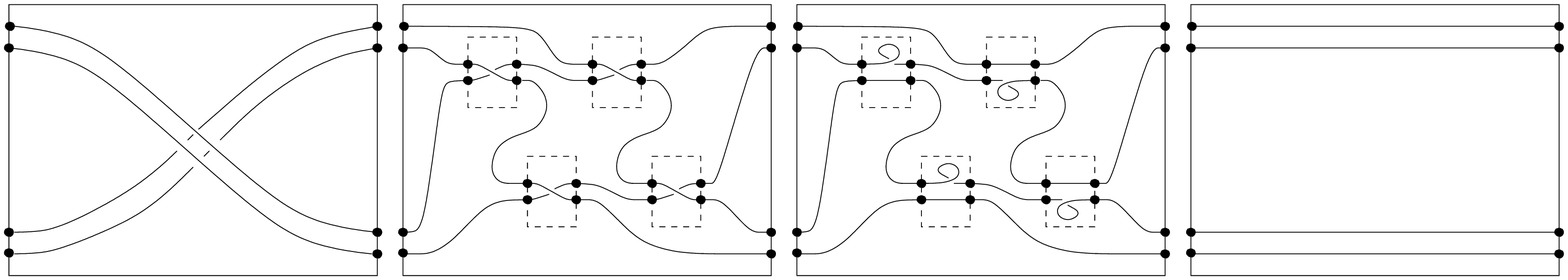,width=6truein}%
\caption{ Exchange of baryon number 2 Skyrmions is contractible in $Q_4$.  }
\label{figexch2}
\end{center}
\end{figure}

Exchange of higher charge Skyrmions may be treated by considering
composites of $B$ unit Skyrmions, as depicted for $B=2$ in figure
\ref{figexch2}. The loop may be deformed into one with four distinct
single exchange events (surrounded by dashed boxes). Each of these may
be replaced by a pair of uncrossed strands, one of which has a $2\pi$
twist, using the homotopy described in figures \ref{fig1} and
\ref{fighwb} in each box. Since each strand has an even number of
twists, this is homotopic to the constant loop. Hence it must be
assigned the phase $(+1)$. The argument clearly generalizes: given an
exchange loop of a pair of charge $B$ composites, one may isolate
$B^2$ single exchange events, each of which can be replaced by a
single twist in one of the uncrossed strands. It is easy to see that
the twists may be distributed so that every strand except at most one
has an even  number of twists. Hence if $B$ is even, this last strand
also has an even  number of twists and the loop is necessarily
contractible. If $B$ is odd, the last strand has an odd number of
twists, so the loop is homotopic to the loop where $2B-1$ Skyrmions
remain static and one Skyrmion executes a $2\pi$ twist. Adding $2B$
anti-Skyrmions, this loop is identified with the baryon number $0$ exchange
loop and hence receives a phase of $(-1)$.

Finkelstein and Rubinstein also model spin in this framework. 
In this model spin is determined by the phase associated to the rotation loop. As we saw in the previous section the rotation and exchange loops agree up to homotopy confirming the spin statistics theorem in this model. This is essentially the
observation that the exchange loop is homotopic to the $2\pi$ rotation
loop (figure \ref{fighwb}).

Note that throughout the above discussion we have used only a local
version of exchange to model particle statistics, and verify the spin-statistics correlation.   This definition of particle statistics and spin makes sense on general $M$, even
without an action of $SO(3)$, because the exchange and rotation loops
have support over a single coordinate  chart, so we have a local
notion of rotating a Skyrmion. Things become much more subtle when a
loop has Pontrjagin-Thom representative which projects to a nontrivial
cycle in $M$. It should not be surprising that it requires a spin
structure on $M$ to specify whether the constituent Skyrmions of such
a loop undergo an even or odd number of rotations.  It was precisely
the generalization from $\R^3$ to general spaces that motivated the
definition of spin structures in the first place. Notice that by
changing the spin structure, we can interpret a loop as either having
an even or odd number of rotations, so one must fix a spin structure
on space before discussing spin. (This is similar to the reason,
discussed in section \ref{scomp} above, why the secondary invariant
for path components of $S^2$-valued maps is only a relative
invariant).  Even in the simple case of quantization of many {\em
point} particles on a topologically nontrivial domain, the statistical
type (boson, fermion or something more exotic) of the particles is
usually taken to be determined only by their exchange  behavior
around trivial loops in $M$ \cite{ptpart}. It may be more reasonable
to require that spin be determined by the behavior of locally
supported rotations, but to insist that the statistical type be
consistent under any particle exchange.  It follows from Proposition
\ref{prop1} that the notion of an exchange or rotation loop around a
contractible loop is well defined independent of the choice of spin
structure. This is just the image of $\pi_4(G)$. That the parity of a
rotation around a non-contractible loop is determined by a spin
structure is explained in Proposition \ref{split} in the next section.

\begin{figure}
\begin{center}
\epsfig{file=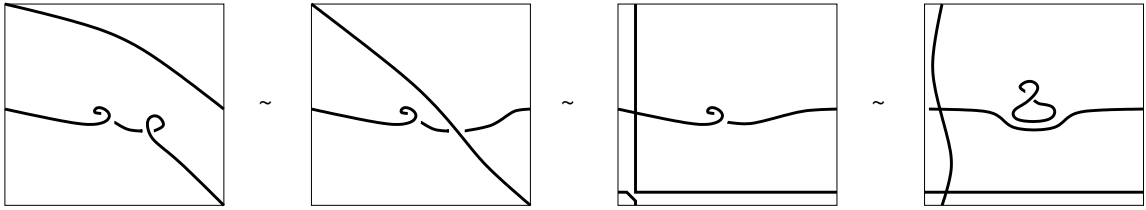,width=6truein}%
\caption{ Sum of a loop with the rotation loop.  }
\label{fig9}
\end{center}
\end{figure}

In fact, the Finkelstein-Rubinstein quantization scheme remains
consistently fermionic in this extended sense provided that the correct representation into U$(1)$ is chosen. As with more traditional models of spin, a spin structure on the domain will be required. When the domain has non trivial first cohomology with $\Z_2$ coefficients there are many spin structures to choose from. Selecting a spin structure produces an isomorphism of  $\pi_1(Q_B)$ with $\Z_2\oplus H_1(M)$.
The required representation is just projection onto the $\Z_2$ factor. Exchange around a (possibly) non-contractible simple closed curve in space means that two identical solitons start at antipodal points on the curve and each one moves without rotating half way around the curve to exchange places with the other soliton. The notion of moving without rotating is where the spin structure enters. We will define this after describing the representation of an exchange displayed in figure 
\ref{fig9}. Each rectangle in this figure represents a slice of a cobordism. The horizontal direction represents time, the vertical direction represents space, and the thick lines are the world lines of the solitons. The top and bottom of each rectangle are to be identified  to make each slice a cylinder representing the curve cross time. We can  imagine different spin structures  obtained by identifying the top and bottom in the straight-forward way or by putting a full twist before making the identification. The first slice is just the exchange around the curve. One of the loops makes a left hand rotation followed by a right hand rotation, but this wobble is the same as no rotation at all. Adding a ribbon between the non-rotating soliton and the right rotation of the bottom soliton produces the second slice. This slice may be described as one soliton making a full left rotation in a fixed location while a second soliton traverses once around the curve without rotation. This slice homotopes to the third slice, and a second ribbon gives the fourth slice. The fourth slice may be described as follows. The vertical S-curve represents the birth of a Skyrmion anti-Skyrmion pair after which the Skyrmion and anti-Skyrmion move in opposite directions around the curve until they collide and annihilate. The horizontal lines are two (nearly) static Skyrmions, and the figure eight curve is a contractible (left) rotation loop. By definition, the exchange is non-rotating with respect to the spin structure if the $\Z_2$ representation of the vertical S-curve resulting from a baryon number $1$ exchange is trivial. We now see that the general exchange is consistent because the two horizontal lines contribute nothing to the representation, the S-curve contributes nothing, and the baryon number $B$ contractible rotation loop contributes $(-1)^B$ as described previously and seen in figure \ref{figexch2}.

As will be seen in section \ref{subsecpc}, there is a close connexion
between $\Sp(1)^M$ and $(S^2)^M$. This allows us to transfer the
Finkelstein-Rubinstein construction of a fermionic quantization scheme
for $\Sp(N)$ valued Skyrmions to the Faddeev-Hopf model. Recall that
here, unless $H^2(M;\Z)=0$, the path  components of $Q$ are not
labelled simply by an integer, but rather are  separated by an
invariant $\alpha\in H^2(M)$, and a relative invariant  $c\in
H^3(M)/2\alpha\cp H^1(M)$. Configurations with $\alpha\neq 0$
necessarily have support which wraps around a nontrivial cycle in
$M$. They therefore lack one of the key point-like characteristics of
conventional solitons: they are not homotopic to arbitrarily highly
localized configurations.  Such configurations are intrinsically tied
to some topological ``defect'' in physical space, and so are somewhat
exotic.  We therefore, mainly restrict our attention to configurations
with $\alpha=0$.  As with Skyrmions, these configurations are labelled
by  $B\in\Z\cong H^3(M;\Z)$ which we identify with the Hopfion
number. This is the relative invariant between the  given
configuration and the constant map. It turns out that all the
$\alpha=0$ sectors, $Q_B$, are homeomorphic (Theorem \ref{fhhom}) and,
further, that $Q_B$ is homotopy equivalent to $(\Sp(1))^M_0$, the
vacuum sector of the classical Skyrme model. The homotopy equivalence
is given by the fibration $f^*$ of Lemma \ref{Fhseq}.  This may be
used to map the charge zero Skyrmion exchange loop to a charge 2
Hopfion exchange loop, generating a $\Z_2$ factor in $\pi_1(Q_2)$. It
follows that Hopfions can be quantized fermionically within the
Finkelstein-Rubinstein scheme. In the more exotic configurations with
$\alpha\neq 0$, the relative invariant takes values in $\Z/(2d\Z)\cong
H^3(M;\Z)/(2\alpha\smile H^1(M;\Z))$. So even though the Hopfion number
is not defined in this case, the parity of the Hopfion number is well
defined and the Finkelstein-Rubinstein formalism still yields a consistently
fermionic quantization scheme.

We return now to the Skyrme model, but in the case where $G$ is not
$\Sp(N)$ but is rather $\Su(N)$, $\Spin$ or exceptional. In this case
the Skyrmion exchange loop is  contractible, so must be assigned phase
$(+1)$ in the Finkelstein-Rubinstein quantization scheme so that only
bosonic quantization is possible in that framework. To proceed, one
may take the wavefunction to be a section of a complex line bundle
over $Q_B$ equipped with a unitary connexion.  Parallel transport with
respect to the connexion associates phases to closed loops in $Q_B$ in
a way that one might hope will mimic fermionic behaviour. The problem
with this is that the holonomy of a loop is not (for non-flat
connexions) homotopy invariant, so the phase assigned to an exchange
loop will depend on the fine detail of how the exchange is
transacted. To get round this, Sorkin introduced a purely topological
definition of statistical type and spin for solitons defined on
$\R^3$. His definition extends immediately to solitons defined on
arbitrary  domains. We review these definitions next.

Recall Sorkin's definition of $f_{\rm stat}:\RP^2\ra Q_2$: choose a
sphere in $\R^3$ and associate to each antipodal pair of points on
this sphere the charge 2 configuration with Pontrjagin-Thom
representation given by that pair of points, framed by the coordinate
basis vectors. The subscript stat in $f_{\rm stat}$ refers to
statistics.  There is an associated homomorphism $f_{\rm
stat}^*:H^2(Q_2)\ra H^2(\RP^2)\cong\Z_2$ given by pullback. According to
Sorkin's definition,  the quantization wherein the wavefunction is a
section of the line bundle over $Q$ associated with class $c\in
H^2(Q;\Z)$ is fermionic if $f_{\rm stat}^*(c)=1$, bosonic otherwise,
\cite{sorkin}. Thinking of $f_{\rm stat}$ as an inclusion map, the
pulled-back class  $f_{\rm stat}^*(c)$ represents the Chern class of
the restriction of the  bundle associated to $c$ over $Q_2$ to the
subset $\RP^2$. The intuition behind this definition is that if there
was a unitary connection on the bundle with parallel transport equal
to $(-1)$ around exchange loops, then the restriction to the bundle to
such a $\RP^2$ would have to be non-trivial. This definition
generalizes to solitons defined on arbitrary domains by analogous maps
from $\RP^2$ into the configuration space based on embeddings of $S^2$
in the domain. To make sense of the framing, one must pick a
trivialization of the tangent bundle of the domain restricted to the
$S^2$. Up to homotopy there is a unique such framing.  The elementary
Sorkin maps will be the ones associated with sufficiently small
spheres that lie in a single coordinate chart.

To model spin for solitons with domain $\R^3$, Sorkin considers the
action of SO$(3)$ on the configuration space given by precomposition
with any field. The orbit of a basic soliton with Pontrjagin-Thom
representative given by one point and an arbitrary frame has
representatives obtained by rotates of the frame. This rotation may be
performed on any isolated lump in any component of configuration space
to define a map $f_{\rm spin}:\hbox{SO}(3)\to Q_B$. Sorkin defines the
quantization associated to a class $c\in H^2(Q_B;\Z)$ to be spinorial
if and only if the pull-back $f_{\rm spin}^*c\in
H^2(\hbox{SO}(3);\Z)\cong \Z_2$ is non-trivial, \cite{sorkin}. This definition generalizes immediately to solitons on arbitrary domains. The
intuition behind this definition is clearly explained in the paper of
Ramadas, \cite{ramadas}. The idea is that the classical SO$(3)$
symmetry of $Q_B$ lifts to a quantum
Sp$(1)=\hbox{SU}(2)=\hbox{Spin}(3)$ symmetry that descends to an
SO$(3)$ action on the space of quantum states if and only if the
pull-back $f_{\rm spin}^*c\in H^2(\hbox{SO}(3);\Z)$ is trivial. In the
case where the domain is not $\R^3$, there is no SO$(3)$ symmetry, but
there is still a SO$(3)$ orbit of any single soliton, obtained by
rotating the framing of a single-point Pontrjagin-Thom representative,
so one still has a local notion of what it means to rotate a soliton.
One may still define a map $f_{\rm spin}:\hbox{SO}(3)\to Q_B$ and
define the quantization corresponding to class $c$ to be spinorial if
and only if $f_{\rm spin}^*c\neq 0$, though there is no corresponding
statement about quantum symmetries.  This should be contrasted with
the case of isospin, which we discuss later in this section. 

Sorkin proves a version of the spin statistics theorem when the domain
is $\R^3$. Recall that rotations may be represented by vectors along
the axis of the rotation with magnitude equal to the angle of
rotation. A one-half rotation in one direction is equivalent to a
one-half rotation in the opposite direction. This gives a natural
inclusion of $\RP^2$ into SO$(3)$ as the set of one-half
rotations. This inclusion induces an isomorphism on cohomology,
$\iota^*:H^2(\hbox{SO}(3);\Z)\to H^2(\RP^2;\Z)$. Sorkin's version of
the spin statistics theorem states that $\iota^*f_{\rm
spin}^*=f_{\rm stat}^*$. There is one slightly stronger version of the
spin statistics correspondence that one may hope for when the domain
is arbitrary. We will discuss this later in this section.

Ramadas proved that the Sorkin definition of statistical type and
spinoriality were strict generalizations of the Finkelstein-Rubinstein
definition when the target group is SU$(N)$. The statement works as
follows. One first notices that the universal coefficient theorem
gives an isomorphism  
$$ 
H^2(Q;\Z)\cong
\hbox{Hom}(H_2(Q;\Z),\Z)\oplus \hbox{Ext}^1(H_1(Q;\Z),\Z) \cong
\hbox{Hom}(H_2(Q;\Z),\Z)\oplus \hbox{Ext}^1(\pi_1(Q),\Z).  
$$ 
When
fermionic quantization is possible in the framework of
Finkelstein-Rubinstein, the exchange loop is an element of order $2$
in $\pi_1(Q)$. Ramadas shows that the corresponding element of
$H^2(Q;\Z)$ pulls back to the non-trivial element of
$H^2(\hbox{SO}(3);\Z)$ under $f_{\rm spin}$ (when the target is
SU$(2)$ so that $f_{\rm spin}$ is defined). More precisely, he shows
several things. He shows that $H^2(\QS{N};\Z)\cong \Z$  for $N>2$  and
$H^2(\QS{2};\Z)\cong \Z_2$.   He shows that the inclusion
SU$(N)\hookrightarrow \hbox{SU}(N+1)$ induces an isomorphism  
$$
H^2(\QS{N+1};\Z)\to H^2(\QS{N};\Z) 
$$ 
for $N>2$ and a surjection
for $N=2$. The $N>2$ case follows from the fibration
SU$(N)\hookrightarrow \hbox{SU}(N+1)\to S^{2N+1}$. The $N=2$ case
follows from the four term exact sequence induced by the Ext functor
together with several ingeniously defined maps, see
\cite{ramadas}. Since $H^2(\QS{2};\Z)\cong \Z_2$, the exchange loop in
$\pi_1(\QS{2})$ corresponds to the generator of $H^2(\QS{2};\Z)$ under
the universal coefficient isomorphism. This class pulls back to the
generator of $H^2(\hbox{SO}(3);\Z)$ under $f_{\rm spin}$. Thus, when
it is possible to quantize fermionically in the
Finkelstein-Rubinstein framework, the exchange loop is an element of
order $2$, so it corresponds to a cohomology class which pulls back
non-trivially under $f_{\rm spin}$ and $f_{\rm stat}$. Hence it is
possible to quantize fermionically in the Sorkin framework, also.

Now turn to the case of an arbitrary domain and compact, simply
connected, simple target group. Given an arbitrary domain, $M$, we can
construct a degree one map to $S^3$ by collapsing the
$2$-skeleton. This map induces, via precomposition, a map between the
corresponding configuration spaces, $\iota_M:G^{S^3}\to G^M$. (Given a
soliton configuration on $S^3$ define one on $M$ by mapping points in
$M$ to points in $S^3$ then follow the configuration into the
target). The induced map on cohomology is surjective, so there is a
portion of the quantization ambiguity ($H^2(Q_B;\Z)$) that depends
only on the codomain and is completely independent of the domain. This
confirms our first physical conclusion, C1. Recall that we call
$H^2(Q_B;\Z)$ the quantization ambiguity because line bundles are
classified by elements of $H^2(Q_B;\Z)$ and wave functions are
sections of such bundles.  By Theorems \ref{thm1} and \ref{thm1co}
this cohomology is   
$$
H^2(Q_B,\Z)=\Z^{b_2(Q_B)}\oplus\tor(H_1(M))\oplus \pi_4(G)  
$$ 
where
$b_2(Q_B)=b_1(M)+1$ for $G=\Su(N)$, $N\geq 3$, and $b_2(Q_B)=b_1(M)$
for $G=\Spin(N)$, $N\geq 7$, $G=\hbox{Sp}(N)$, $N\geq 1$, or $G$
exceptional. Here $b_k(X)$ denotes the  $k^{th}$ Betti number of $X$,
that is, $\dim_\R\, H_k(X;\R)$. Also $\pi_4(G)=\Z_2$ for
$G=\hbox{Sp}(N)$, $N\geq 1$, and $\pi_4(G)=0$ otherwise.  Notice that
the elementary Sorkin maps factor through $\iota_M:G^{S^3}\to G^M$.
Use $f_{\rm Estat}$ and $f_{\rm Espin}$ to denote the elementary
Sorkin maps defined on an arbitrary domain. By definition we have,
$f_{\rm Estat}=\iota_M\circ f_{\rm stat}$ and $f_{\rm
Espin}=\iota_M\circ f_{\rm spin}$.  If $G=\Spin(N)$,  $N\geq 7$, or
exceptional, then $H^2(G^{S^3};\Z)=0$ so $\iota_M^*$ is
trivial. Hence, fermionic quantization is impossible in these
cases. The inclusions Sp$(N)\hookrightarrow\hbox{Sp}(N+1)$ induce maps
on the configuration spaces $\Sp(n)^{S^3}$ that induce isomorphisms on
cohomology and Sp$(1)\cong\Su(2)$, so we have reduced to the special
unitary case.  If $G=\Su(N)$, $N\geq 3$, then $H^2(G^{S^3};\Z)=\Z$ and
$\iota_M^*$  maps $\tor(H_1(M))$ and all the generators
$\mu(\Sigma_1^k\otimes x_3)$ to $0$, and $\mu([M]\otimes x_5)$ to
$\mu([S^3]\otimes x_5)$.  Since the map on cohomology induced by
$\iota_M$ is surjective, fermionic quantization in the generalization
of the sense of Sorkin is possible over an arbitrary domain if and
only if it is possible for domain $S^3$. Combined with the result of
Ramadas that $f_{\rm spin}:H^2(G^{S^3};\Z)\to H^2(\hbox{SO}(3);\Z)$
surjects and Sorkin's spin statistics correlation that $f_{\rm
stat}^*=\iota^*f_{\rm spin}$ one obtains $f_{\rm stat}(m\mu([M]\otimes
x_5))= m\in\Z_2$, so quantization on the bundle represented by the
class $m\mu([M]\otimes x_5)$ is fermionic if and only if $m$ is odd.
Our second physical conclusion, C2, establishing necessary and
sufficient conditions for the existence of fermionic quantizations
follows from these comments.

There are consistency conditions that one would like to check with
regard to the generalized Sorkin model of particle statistics.  Since
$Q_B$ is connected and $H^2(Q_0,\Z)$ is discrete, there is a canonical
isomorphism $H^2(Q_0)\ra H^2(Q_B)$, so a class $c\in H^2(Q_0;\Z)$
defines a fixed class over each sector $Q_B$. As in the discussion of
the Finkelstein-Rubinstein model of particle statistics, we would like
to know that Baryon number $B^\prime$ lumps in $Q_B$ are all bosonic
when the one lump class in $Q_1$ is bosonic ($f_{\rm Estat}^*c=0$).
We would also like to know that Baryon number $B^\prime$ lumps in
$Q_B$ are bosonic or fermionic according to the parity of $B^\prime$
when the one lump class in $Q_1$ is fermionic ($f_{\rm
Estat}^*c=1$). The statistical type of a Baryon number $B^\prime$ lump
may be defined via a generalization of the Sorkin map in which a pair
of Baryon number $B^\prime$ lumps is placed at antipodal points on a
sphere. With this definition, cobordism arguments similar to those
given in the Finkelstein-Rubinstein case show that this model of
particle statistics is indeed consistent. 

As in the Finkelstein-Rubinstein case, the spin of a particle will
just be determined by a local picture, and the statistical type may be
based on non-local exchanges of identical particles. A non-local
exchange will be defined by a generalization of the Sorkin map
associated to an arbitrary embedded $2$-sphere, say $S\hookrightarrow
M$. Denote the associated map by, $f_{\rm S:stat}:\RP^2\ra Q_B$. There
are two cases: either $S$ separates the domain, so $M-S=M_1\cup M_2$,
or $S$ does not separate.  If the 2-sphere in $M$ separates, we can
define a degree one map $p:M\to S^3$ by collapsing the relative
$2$-skeleta of $\overline M_1$ and $\overline M_1$. As in the case
of the elementary Sorkin map, we obtain $f_{\rm S:stat}=f_{\rm
stat}\circ \widehat p$ where $\widehat p:G^{S^3}\to G^M$. It follows
that this model of particle statistics is consistent with these
non-local exchanges. If the sphere does not separate, then there is a
simple path from one side of the sphere to the other side of the
sphere. A tubular neighborhood of the union of this path and the
sphere is homeomorphic to a punctured $S^2\times S^1$. We may
construct a degree one projection from $M$ to $S^2\times S^1$ by
collapsing the complement of this tubular neighborhood. This
intertwines the Sorkin map defined using the non-separating sphere
with the Sorkin map defined using $S^2\times\{1\}\subset S^2\times
S^1$. If we knew the following conjecture, then this model of particle
statistics would be consistent in this larger sense. As it is, we know
that it satisfies the stronger consistency condition in the typical
case where the domain does not contain a non-separating sphere.

\noindent
{\bf Conjecture} 
$$
f^*_{S^2\times \{1\}:{\rm stat}}\mu([S^2\times S^1]\otimes x^5)\neq 0.
$$  

To discuss isospin, we recall some standard facts about extending
group actions on a configuration space. Given a complex line bundle over
$Q$, let $\widehat Q$ be the associated principal U$(1)$ bundle. If a group
$\Gamma$ acts on $Q$ It is possible to construct an extension of $\Gamma$ 
by U$(1)$
that acts on $\widehat Q$ so that the projection to $\Gamma$
intertwines the two actions. The extension $\widehat \Gamma$ may be defined
as equivalence classes of paths in $\Gamma$, see \cite{bms-sp}. The quantum 
symmetry group is a subgroup of this extension. When
$\Gamma=\hbox{SO}(3)$ the possible U$(1)$ extensions are
SO$(3)\times\hbox{U}(1)$ and U$(2)$. These correspond to integral and
fractional isospin respectively when SO$(3)$ acts as rotations on the
target. Recall that every compact, simply connected, simple Lie group 
$G$ has a $\Sp(1)$ subgroup. We define the isospin action on $G$ to be
the adjoint action of this $\Sp(1)$ subgroup. This coincides with the
usual definition if $G=\Sp(1)$. Of course, we can always take 
a trivial line bundle over $Q$, so any of our configuration spaces admit
quantizations with integral isospin, confirming our fifth physical 
conclusion, C5.

To justify our remaining physical conclusions about isospin, we review the 
required
constructions.  The Sorkin map SO$(3)\to \hbox{SU}(2)^{S^3}$
is the map obtained by the isospin action. To see this, notice that we can 
rotate the frame in the Pontrjagin-Thom representative by either rotating 
the domain or by rotating the codomain. When a class in
$H^2(G^M;\Z)$ pulls back to the generator of $H^2(\RP^2;\Z)$
under the Sorkin map, we claim that the associated quantization 
has fractional isospin. Assume otherwise so the extension of the
rotation group is SO$(3)\times\hbox{U}(1)$. This means that the
SO$(3)$ subgroup is a lift of the SO$(3)$ action on the configuration
space to the bundle over the space. Restricting this to the image of SO$(3)$ 
under $f_{\rm spin}$ we
obtain a contradiction from Theorem \ref{thmgot}. 
Since such classes exist whenever the configuration space admits a fermionic 
quantization, we obtain our third physical conclusion, C3.
To show that quantizations with fractional isospin are not possible when the 
group does not have a symplectic or special unitary factor, one must just 
follow through the construction of the extension given in  \cite{bms-sp} to 
see that the resulting extension is trivial. This establishes our fourth 
conclusion, C4.

As we noted earlier, the relation between the configuration space of 
Sp$(1)$-valued maps and $S^2$-valued maps implies that it is always possible 
to fermionically quantize Hopfions. Since it is possible to quantize 
Sp$(1)$-valued solitons with fractional isospin, the same relation implies 
that it is possible to quantize $S^2$-valued solitons with fractional 
isospin. This is our sixth physical conclusion, C6.

\section{Proofs}
\label{proofs}
\label{pro}
\news

We begin by recalling some basic homotopy and homology  theory
\cite{span}. For pointed spaces $X,Y$, let $[X,Y]$ denote the set of
based homotopy classes of maps $X\ra Y$. There is a distinguished
element $0$ in $[X,Y]$, namely the class of the constant  map. Given a
map $f:X\ra X'$ there is for each $Y$ a natural map $f^*:[X',Y]\ra
[X,Y]$ defined by composition. We define $\ker f^*\subset [X',Y]$ to
be the  inverse image of the null class $0\in[X,Y]$. A sequence of
maps  $X\stackrel{f}{\ra}X'\stackrel{g}{\ra}X''$ is {\em coexact} if
$\ker f^*=\im g^*$ for every choice of codomain $Y$. Longer sequences
of maps are coexact if every constituent triple is coexact. Note that
this makes sense even in the absence of group structure. If $Y$
happens to be a Lie group $G$, as it will be for us, then $[X,G]$
inherits a group structure by pointwise multiplication, $f^*$ and
$g^*$ are homomorphisms, and the sequence
$[X'',G]\stackrel{g^*}{\ra}[X',G]\stackrel{f^*}{\ra}[X,G]$ is exact in
the usual sense. In the following, we will make extensive use of the
following standard result \cite{span}:

\begin{prop}\label{coexact} If $X$ is a CW complex and $A\subset X$ is a 
subcomplex then there is an infinite coexact sequence,
$$
A\hookrightarrow X\ra X/A\ra SA\hookrightarrow SX\ra S(X/A)\ra\cdots
\ra S^nA\hookrightarrow S^nX\ra S^n(X/A)\ra\cdots
$$
where $S^n$ denotes iterated suspension.
\end{prop}

The proofs will use several naturally defined homomorphisms. Any map
$f:X\ra Y$ defines homomorphisms $f_*:H_k(X)\ra H_k(Y)$ which depend
on $f$ only up to homotopy. Hence, one has natural maps ${\cal
H}_k:[X,Y]\ra \homo{H_k(X)}{H_k(Y)}$. 
 There is a natural (Hurewicz) homomorphism $\hur_k:\pi_k(X)\ra
H_k(X)$ sending each map $S^k\ra X$ to the push-forward of the
fundamental class via the map in $X$. If $X$ is $(k-1)$-connected then
$\hur_k$ is an isomorphism.  There is also a natural isomorphism
$\susp_k:H_k(SX)\ra H_{k-1}(X)$ relating the homologies of $X$ and
$SX$.

We may now prove a preliminary lemma that is used in the computation
of both the fundamental group and the real cohomology. This lemma is
the place where  we use the assumption that the domain is
orientable. This lemma was used in  \cite{AK1} as well. Note that
$SX^{(k)}$ denotes the suspension of the  $k$-skeleton of $X$. The
$k$-skeleton of the suspension of $X$ will always be denoted
$(SX)^{(k)}$.

\begin{lemma}\label{surj}
For a closed, connected, orientable $3$-manifold, and
simply-connected,  compact Lie group the map,
$$
[SM,G]\to [SM^{(2)},G]
$$
induced by inclusion is surjective.
\end{lemma}
{\it Proof:\, } Start with a cell decomposition of $M$ with exactly
one $0$-cell and exactly  one $3$-cell. The sequences,
\begin{diagram}
M/M^{(2)} & \rTo^{\partial} & SM^{(2)} & \rTo & SM, 
\end{diagram}
\vskip-.1in
\noindent and
\vskip-.1in
\begin{diagram}  
SM^{(1)} & \rTo &  SM^{(2)} & \rTo{q} & S(M^{(2)}/M^{(1)}),
\end{diagram}
are coexact by Proposition \ref{coexact} with $X=M$, $A=M^{(2)}$ and
$X=M^{(2)}$, $A=M^{(1)}$ respectively. Hence, the sequences, 
\begin{diagram}
[SM,\, G]&\rTo&[SM^{(2)},\, G]&\rTo^{\cd^*}&[M/M^{(2)},\, G], 
\end{diagram}
\vskip-.1in
\noindent and
\vskip-.1in
\begin{diagram}
[S(M^{(2)}/M^{(1)}),\, G]&\rTo^{q^*}&[SM^{(2)},\, G]&\rTo&[SM^{(1)},\,
G]=0,
\end{diagram}
are exact. The group $[SM^{(1)},\, G]$ is trivial because $G$ is
$2$-connected. Hence $q^*$ is surjective.  The space
$\,M^{(3)}/M^{(2)}\,$ is homeomorphic to $\,D^3/S^2$. Under this
identification, $\,\partial(x) = \left(f^{(3)}(\frac{x}{|x|}),
|x|\right)$  for $\,x\in D^3$, where $\,f^{(3)}:\,S^2\to M^{(2)}\,$ is
the attaching map  for the 3-cell.  Now we can construct the following
commutative diagram:
\begin{diagram}
[S(M^{(2)}/M^{(1)}), G] & \rTo^{\partial^*\circ q^*} &
[M^{(3)}/M^{(2)}, G]\\ \dTo^{{\cal H}_3} & & \dTo_{{\cal H}_3}\\
\homo{H_3(S(M^{(2)}/M^{(1)}))}{H_3(G)}&
&\homo{H_3(M^{(3)}/M^{(2)})}{H_3(G)}\\ \dTo^{\susp_3^{-1\,
*}\circ\hur_{3\, *}}& & \dTo_{\hur_{3\, *}}\\
\homo{H_2(M^{(2)}/M^{(1)})}{\pi_3(G)}&\rTo^{\delta}&
\homo{H_3(M^{(3)}/M^{(2)})}{\pi_3(G)}.
\end{diagram}
Note that both maps ${\cal H}_3$ are isomorphisms, and 
$\hur_3:\pi_3(G)\ra H_3(G)$
is an isomorphism because $G$ is 2-connected, so the vertical maps are
all isomorphisms. We may therefore identify $\cd^*\circ q^*$ with
the map $\delta$. But $\delta$ is the coboundary map in the CW cochain
complex for $H^*(M;\pi_3(G))$. Since $M$ is orientable, this coboundary is 
trivial. Since $q^*$ is surjective, we conclude that $\partial^*=0$. \hfill
$\Box$

\subsection{The fundamental group of the Skyrme configuration space}
\label{subsecpa}
We saw in section \ref{hom} that it was sufficient to study  simple,
simply-connected Lie groups. We begin our study of  the fundamental
group by showing that the fundamental group of the  configuration
space fits into a short exact sequence in this case.

\begin{prop}\label{prop1}
If $M$ is a closed, connected, orientable $3$-manifold and $G$ is a
simple,  simply-connected Lie group, then
$$
0\to \pi_4(G)\to\pi_1(G^M)\to H^2(M;\pi_3(G))\to 0
$$
is an exact sequence of abelian groups. The maps in this sequence will
be defined in the course of the proof.
\end{prop}
{\it Proof:\, } We have $\pi_1(G^M)=[S^1,\, G^M]\cong [SM,\, G]$,
where $SM$ is the suspension of $M$. The sequence
$$
(SM)^{(3)}\hookrightarrow SM \to SM/(SM)^{(3)}
$$
is coexact by Proposition \ref{coexact} with $X=SM$, $A=(SM)^{(3)}$,
and hence induces the exact sequence of groups:
\begin{equation}\label{sq1}
[SM/(SM)^{(3)},\, G]\to [SM,\, G]\to [(SM)^{(3)},\, G].
\end{equation}
Noting that $(SM)^{(3)}=SM^{(2)}$, Lemma \ref{surj} implies that the
last map  in the above sequence is surjective.  This will become the
exact sequence we seek, after suitable identifications.  First we
identify the third group in  sequence \ref{sq1} with the second
cohomology of $M$, in similar fashion to the proof of Lemma
\ref{surj}. From Proposition \ref{coexact}, we have exact  sequences,
\begin{diagram}
[S((SM)^{(2)}/(SM)^{(1)}),\, G] &  \rTo^{q^*} & [S((SM)^{(2)}),\, G] &
\rTo & [S((SM)^{(1)}),\, G] 
\end{diagram}
\vskip-.1in

\noindent
($X=(SM)^{(2)}$, $A=(SM)^{(1)}$) and

\vskip-.1in
\begin{diagram}  
[S((SM)^{(2)}),\, G] & \rTo{\partial^*} &   [(SM)^{(3)}/(SM)^{(2)},\,
G] & \rTo &  [(SM)^{(3)},\, G],
\end{diagram}
($X=(SM)^{(3)}$, $A=(SM)^{(2)}$). Since $G$ is $2$-connected, 
$[S((SM)^{(1)}),\,G]=0$, and $q^*$ is surjective.  
Thus coker$(\partial^*\circ
q^*)=\hbox{coker}(\partial^*)=[(SM)^{(3)},\, G]$.  Using the Hurewicz
and suspension isomorphisms as before, we may identify
$\partial^*\circ q^*$ with a  coboundary map in the CW cochain complex
for $H^*(M;\pi_3(G))$:
\begin{diagram}
[S((SM)^{(2)}/(SM)^{(1)}),\, G] & \rTo^{\partial^*\circ q^*} &
[(SM)^{(3)}/(SM)^{(2)},\, G]\\ \dTo & & \dTo \\
\hbox{Hom}(H_2((SM)^{(2)}/(SM)^{(1)}),\pi_3(G)) & 
\hskip-3pt\rTo_{\delta}\hskip-3pt &
\hbox{Hom}(H_3((SM)^{(3)}/(SM)^{(2)}),\pi_3(G)). \\ 
\end{diagram}
Now,
$$
[(SM)^{(3)},G] \cong\hbox{coker}(\partial^*\circ q^*)
\cong\hbox{coker}(\delta) \cong H^3(SM;\pi_3(G))\cong H^2(M;\pi_3(G)).
$$

The first group in sequence \ref{sq1} is $\pi_4(G)$ because
$SM/(SM)^{(3)}\cong S^4$. For the non-symplectic groups $\pi_4(G)=0$,
and we  are done. For the higher symplectic groups, the fibration
$$
\Sp(n)\to \Sp(n+1)\to S^{4n+3}
$$
induces a fibration
$$
(\Sp(n))^M\to (\Sp(n+1))^M\to (S^{4n+3})^M.
$$
The homotopy exact sequence of this fibration reads
$$
\pi_2((S^{4n+3})^M)\to\pi_1((\Sp(n))^M)\to \pi_1((\Sp(n+1))^M)\to
\pi_1((S^{4n+3})^M).
$$
Now $\pi_k((S^{4n+3})^M)=[S^k,\, (S^{4n+3})^M] \cong [S^kM,\,
S^{4n+3}]$. For  $k=1, 2$ these groups are trivial since $S^{4n+3}$ is
$5$-connected.  Hence $\pi_1({\Sp}(n+1)^M)\cong \pi_1({\Sp}(n)^M)$ for
all  $n\geq 1$, so the proposition reduces to showing that the first
map in  sequence \ref{sq1} is injective for $G=\Sp(1)$. In the special
case of  $M\cong S^3$ the exchange loop depicted in
figure \ref{fig1} represents the generator of
$\pi_4(\hbox{Sp}(1))\cong \pi_1((\hbox{Sp}(1))^{S^3})$. Our final task
is to  show that the image of this generator under push forward by the
collapsing map $M\ra M/M^{(2)}$ is non-trivial in
$\pi_1((\Sp(1))^{M})$. 

Proceed indirectly and assume that there is a homotopy between the
constant  loop and the exchange loop, say $H:M\times[0,1]\times [0,1]
\to \hbox{Sp}(1)$.  Set $\Sigma=H^{-1}(-1)$. Now glue the homotopy
from figure \ref{fig3} to  this homotopy and let
$\widehat\Sigma=\Sigma\cap \Delta$ where $\Delta$ is  the hemisphere
in figure \ref{fig3}. The trivialization of the normal bundle  to
$\widehat\Sigma$ defined over $\Sigma$ and the trivialization of the
normal bundle to $\widehat\Sigma$ defined over $\Delta$ do not
match. The discrepency is the generator of $\pi_1(\hbox{SO}(3))$. It
follows that the second  Stiefel-Whitney class,
$w_2(N(\widehat\Sigma))\in  H^2(M\times[0,1]\times [0,1];
\pi_1(\hbox{SO}(3)))$ is non-trivial \cite{ste}.  However, the Whitney
product formula yields, 
$$
\begin{array}{rl}
w(N(\widehat\Sigma))= & w(N(\widehat\Sigma))\cp w(T\widehat\Sigma)\\ =
& w(T\widehat\Sigma\oplus N(\widehat\Sigma)) =  w(T(M\times[0,1]\times
[0,1])|_{\widehat\Sigma}) = 1. 
\end{array}
$$
Here $w$ is the total Stiefel-Whitney class, $ w(T\widehat\Sigma)=1$
because  the Stiefel-Whitney class of any orientable surface is $1$,
and  $T(M\times[0,1]\times [0,1])|_{\widehat\Sigma}$ is trivial. This
contradiction establishes the proposition.  \hfill$\Box$

It is well known that a split exact sequence of abelian groups, $0\to
K\to G  \stackrel{\leftarrow}{\rightarrow} H \to 0$, induces an
isomorphism,  $K\oplus H \cong G$. The following proposition will
establish such a splitting, and therefore, complete our  computation
of the fundamental group of the Skyrme configuration spaces. The proof
will require surgery descriptions of $3$-manifolds, so we recall what
this means. Given a framed link, say $L$, (i.e. $1$-dimensional
submanifold with trivialized normal bundle or identification of a
closed tubular neighborhood with $\disju S^1\times D^2$) in
$S^3=\partial D^4$, we define a $4$-manifold by $D^4\cup_{\disju
S^1\times D^2}D^2\times D^2$. The boundary of this $4$-manifold is
said to be the $3$-manifold obtained by surgery on $L$. It is denoted,
$S^3_L$.

\begin{figure}
\hskip105bp\epsfig{file=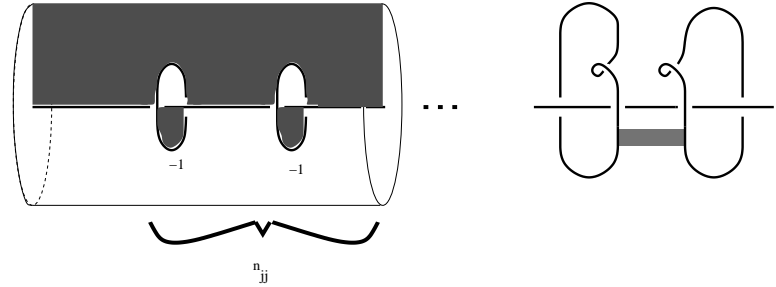,width=4truein}%
\caption{The $2$-cycle in the proof of Proposition \ref{split}}
\label{fig4}
\end{figure}

\begin{prop}\label{split}
The sequence,
$$
0\to \pi_4(G)\to\pi_1(G^M)\to H^2(M;\pi_3(G))\to 0
$$
splits, and there is a splitting associated to each spin structure on
$M$.
\end{prop}
{\it Proof:\, } As we saw at the end of the proof of the previous
proposition, it is  sufficient to check the result for
$G=\hbox{Sp}(1)$. Since the three  dimensional Spin cobordism group is
trivial, every $3$-manifold is surgery  on a framed link with even
self-linking numbers \cite{kirby}.  
 Such a surgery description induces a Spin structure in $M$. Let
$M=S^3_L$ be  such a surgery description, orient the link and let
$\{\mu_j\}_{j=1}^c$ be the positively oriented meridians to the
components of the link. These meridians  generate $H_1(M)\cong
H^2(M;\pi_3(\hbox{Sp}(1)))$. This last isomorphism is  Poincar\'e
duality. Define a splitting by:
$$
s:H_1(M)\to\pi_1((\hbox{Sp}(1))^M); \quad s(\mu_j)=
PT(\mu_j\times\{\frac12\}, \hbox{canonical framing}).
$$
Here $PT$ represents the Pontrjagin-Thom construction and the
canonical  framing is constructed as follows. The first vector is
chosen to be the  $0$-framing on $\mu_j$ considered as an unknot in
$S^3$. The second vector is  obtained by taking the cross product of
the tangent vector with the first  vector, and the third vector is
just the direction of the interval. We will  now check that this map
respects the relations in $H_1(M)$. Let  $Q_L=(n_{jk})$ be the linking
matrix so that,  $H_1(M)=\langle \mu_j| n_{jk}\mu_j=0 \rangle$. We are
using the summation  convention in this description. The $2$-cycle
representing the relation,  $n_{jk}\mu_j=0$ may be constructed from a
Seifert surface to the  $j^{\rm th}$ component of the link, when
this component is viewed  as a knot in $S^3$. Let $\Sigma_j$ denote
this Seifert surface. The desired  $2$-cycle is then
$\widehat\Sigma_j=(\Sigma_j-\stackrel{\circ}{N}(L))\cup\sigma_j$. Here
$\sigma_j$ is the surface in $S^1\times D^2$ with $n_{jj}$ meridians
depicted  on the left in figure \ref{fig4}.
The boundary of $\widehat\Sigma_j$ is exactly the relation,
$n_{jk}\mu_j=0$.  The framing on each copy of $\mu_k$ for $k\neq j$
induced from this surface  agrees with the $0$-framing. The framing on
each copy of $\mu_j$ is  $-\hbox{sign}(n_{jj})$. The surface,
$\widehat\Sigma_j$ may be extended to a  surface in $M\times
[0,1]\times [0,1]$ by adding a collar of the boundary in  the
direction of the second interval followed by one band for each pair of
the $\mu_j$ as depicted on the right of figure \ref{fig4}. The
resulting  surface has a canonical framing, and the corresponding
homotopy given by  the Pontrjagin-Thom construction homotopes the loop
corresponding to the  relation to a loop corresponding to a
$\pm2$-framed unlink. Such a loop is  null-homotopic, as
required. \hfill $\Box$

We remark that the Spin structures on $M$ correspond to
$H^1(M;{\mathbb Z}_2)$.  In addition, the splittings of ${\mathbb
Z}_2\to \pi_1(\hbox{Sp}(1)^M)\to  H^2(M;{\mathbb Z})$ corresponds to
the group cohomology, 
$$
H^1(H^2(M;{\mathbb Z});{\mathbb Z}_2) \cong   H^1(H_1(M;{\mathbb
Z});{\mathbb Z}_2) \cong H^1(M;{\mathbb Z}_2).
$$
The last isomorphism is because the $2$-skeleton of $M$ is the
$2$-skeleton  of a $K(H_1(M;{\mathbb Z}), 1)$. A combination of
Propositions \ref{prop1} and \ref{split} together with Reductions
\ref{r5}, \ref{r6} and \ref{r7} and the corollary  of the universal
coefficient theorem that $H^*(X;A\oplus B)\cong H^*(X;A)\oplus
H^*(X;B)$ give Theorem \ref{thm1}.

\subsection{Cohomology of Skyrme configuration spaces}
\label{subsecpb} 

As we have seen we may restrict our attention to compact, simple,
simply-connected $G$. Recall the cohomology classes, $x_j$,  and the
$\mu$-map defined in section \ref{geo}.  Throughout this section we
will take the coefficients of any homology or  cohomology to be the
real numbers unless noted to the contrary. To compute the  cohomology
of $G^M$ we will use the cofibrations  $M^{(k)}\to M^{(k+1)} \to
M^{(k+1)}/ M^{(k)}$ and the fact that  $ M^{(k+1)}/ M^{(k)}$ is a
bouquet of spheres to reduce the problem to the  case where the domain
is a sphere. 

Briefly recall the computation of the cohomology of the loop
spaces. These are well known results, but we sketch a proof because
this explains why the classes $\mu(\Sigma,x_jx_k)$ are trivial.  As
usual let $\Omega^kG=G^{S^k}$ denote the $k$-iterated loop space. We
have the following lemma.
\begin{lemma}\label{og}
The cohomology rings of the first loop groups are given by,
$H^*(\Omega G)={\mathbb R}[\mu([S^1]\otimes x_j)]$, $H^*(\Omega^2
G)={\mathbb R}[\mu([S^2]\otimes x_j)]$, and $H^*(\Omega^3_0G)={\mathbb
R}[\mu([S^3]\otimes x_j),j>3]$.
\end{lemma}
{\it Proof:\, } Recall that the path space, $PG$ is contractible, and
fits into a fibration, $\Omega G\hookrightarrow PG\to G$. The Serre
spectral sequence of this  fibration has, $E_2^{p,q} =
H^p(G;H^q(\Omega G))$. Since $G$ is  simply-connected, the coefficient
system is untwisted. Since $PG$ is contractible all classes of
positive degree have to die at some point in this spectral
sequence. By location we know that all differentials of $x_3$ vanish,
so there must be some class in $H^2(\Omega G)$ mapping to $x_3$. The
class $\mu([S^1]\otimes x_3)$ is one such class, and the only class
that there can be without having something else live to the limit
group of the spectral sequence. Notice that classes of the form
$x_3\prod x_{j_k}$ are images of classes of the form $\mu([S^1]\otimes
x_3) \prod x_{j_k}$, so we have killed all classes with a factor of
$x_3$. In the same way, we can kill terms with a factor of the next
$x_j$. We  conclude that $H^*(\Omega G)={\mathbb R}[\mu([S^1]\otimes
x_j)]$. Repeating the argument with the
fibration,  $\Omega^2G\to P\Omega G\to \Omega G$ we obtain,
$H^*(\Omega^2 G)={\mathbb R}[\mu([S^2]\otimes x_j)]$. This time the
coefficient system is untwisted because, $\pi_1(\Omega G)\cong
\pi_2(G) = 0$.

We need to adjust the argument a bit at the next stage because
$\pi_0(\Omega^3G)\cong\pi_1(\Omega^2 G)\cong \pi_3(G) \cong {\mathbb
Z}$.  This shows that the path components of $\Omega^3G$ may be
labeled by the  integers. Each component is homeomorphic to the
identity component since  $\Omega^3G$ is a topological group.  In this
case, we have no guarantee that the coefficient system is untwisted,
so we will use a different approach that we be useful again in section
\ref{subsecpd}.  Let $\widetilde{\Omega^2G}$ denote the  universal cover
of $\Omega^2G$ and let $\Omega_0^3G$ denote the identity  component of
$\Omega^3G$.  These fit into a fibration,  $\Omega_0^3G\to
P\widetilde{\Omega^2G} \to \widetilde{\Omega^2G}$ that may be used to obtain,
$H^*(\Omega^3_0G)={\mathbb R}[\mu([S^3]\otimes x_j),j>3]$. We will use 
equivariant cohomology to compute the cohomology of  $\widetilde{\Omega^2G}$.

Recall that any Lie group, say $\Gamma$, acts properly on a
contractible space called the total space of the universal
bundle. This space is denoted $E\Gamma$. The quotient of this by
$\Gamma$ is the classifying space $B\Gamma$. Let $X$ be a $\Gamma$
space (i.e. a space with a $\Gamma$ action) and consider the space
$X_\Gamma :=E\Gamma\times_\Gamma X$. The cohomology of the space
$X_\Gamma$ is called the equivariant cohomology of $X$. It is denoted
by $H^*_\Gamma(X)$. When the $\Gamma$ action on $X$ is free and proper
(as it is in our case), we have a fibration $X_\Gamma\to X/\Gamma$
obtained by ignoring the $E\Gamma$ component in the definition of
$X_\Gamma$. The fiber of this fibration is just $E\Gamma$ which is
contractible, so the spectral sequence of the fibration implies that
the cohomology of $X/\Gamma$ is isomorphic to the equivariant
cohomology of $X$. By ignoring the $X$ component in the definition of
$X_\Gamma$ we obtain a fibration $X_\Gamma\to B\Gamma$ that may be
used to relate the equivariant cohomology of $X$ to the cohomology of
$X$.

If we apply these ideas with $X=\widetilde{\Omega^2G}$ and
$\Gamma=\pi_1(\Omega^2G)\cong \Z$, we obtain a spectral sequence that
may be used to show that the cohomology of $\tilde{\Omega^2G}$ is
generated by  $\mu([S^2]\otimes x_j)$ for $j>3$. This then plugs in to
give the stated result for $\Omega^3_0G$. \hfill $\Box$

Returning to the situation of a $3$-manifold domain, let $M$ have a
cell decomposition with one $0$-cell ($p_0$) several  $1$-cells
($e_r$) several $2$-cells ($f_s$) and one $3$-cell ($[M]$). Since
$G^{X\vee Y}=G^X\times G^Y$, and $ M^{(k+1)}/ M^{(k)}$ is a bouquet of
spheres we have, 
$$
\begin{array}{rcl} H^*(G^{M^{(1)}})&=&{\mathbb R}[\mu(e_r\otimes
x_j)], \nonumber \\ H^*(G^{M^{(2)}/M^{(1)}})&=&{\mathbb
R}[\mu(f_s\otimes x_j)],\nonumber \\
H^*(G^{M^{(3)}/M^{(2)}}_0)&=&{\mathbb R}[\mu([M]\otimes x_j), j>0].
\end{array}
$$ 
The next lemma assembles these facts into the cohomology of
$G^{M^{(2)}}$.

\begin{lemma}\label{gm2}
If $\Sigma^1_r$ form a basis for $H_1(M)$, and $\Sigma^2_s$ form a
basis for  $H_2(M)$ we have,
$$
H^*(G^{M^{(2)}})= {\mathbb R}[\mu(\Sigma^1_r\otimes x_j),
\mu(\Sigma^2_s\otimes x_k)].
$$
\end{lemma}
{\it Proof:\, } The cofibration $M^{(1)}\to M^{(2)} \to M^{(2)}/
M^{(1)} $ leads to a  fibration,
$$
G^{M^{(2)}/ M^{(1)}}\to G^{M^{(2)}}\to G^{M^{(1)}}.
$$ 
Since $G$ is $2$-connected, $\pi_1(G^{M^{(1)}})=0$, so the
coefficients in  the cohomology appearing in the second term of the
Serre spectral sequence  are untwisted. We have,  
$$
\begin{array}{rcl}
E_2^{*,*}&=&H^*( G^{M^{(1)}};  H^*(G^{M^{(2)}/M^{(1)}}))  \\
&=& H^*(G^{M^{(1)}})\otimes  H^*(G^{M^{(2)}/M^{(1)}})  \\
&\cong&  {\mathbb R}[\mu(e_r\otimes x_j), \mu(f_s\otimes x_k)]. 
\end{array}
$$
To go further we need to understand the differentials in this spectral
sequence. Since $\mu(e_r\otimes x_j)\in E_2^{j-1,0}$ we have
$d_k\mu(e_r\otimes x_j)=0$ for all $k$. We will show that  $d_\ell
\mu(f_s\otimes x_k)=0$ for  $\ell<k-1$ and $d_{k-1} \mu(f_s\otimes
x_k)=-\mu((\partial f_s)\otimes x_k)$ from which the result will
follow. The multiplication on $G$ induces a ring  structure on the
homology of $G^{M^{(1)}}$ and $G^{M^{(2)}/M^{(1)}}$. Using  the
homology spectral sequence of the path fibrations, one may show that
these homology groups are generated by cycles
$e_r\beta_j:\Sigma_r^{j-1}\to G^{M^{(1)}}$ and
$f_s\beta_k:\Sigma_s^{k-2}\to G^{M^{(2)}/M^{(1)}}$ dual to
$\mu(e_r\otimes x_j)$ and $\mu(f_s\otimes x_k)$ respectively. The
product in  the homology ring of $G^X$ is given by
$\beta\cdot\beta^\prime:\Sigma\times\Sigma^\prime\to G^X$ with
$\beta\cdot\beta^\prime(x, y)(p)=\beta(x)(p)\beta^\prime(y)(p)$. The
computation \ref{delta} in section \ref{geo} shows that the differential 
of our spectral sequence is given by 
$$
(d_\ell \mu(f_s\otimes x_k))(\beta\otimes\beta^\prime)=-
\int_{(\partial f_s)\times\Sigma\times\Sigma^\prime}  w^*x_k,
$$
where $w(p, x, y)= \beta(x)(p)\beta^\prime(y)(\pi(p))$ and
$\pi:M^{(2)}\to M^{(2)}/M^{(1)}$ is the canonical projection. We are
using  the integral as a suggestive notation for the cap product. We
see that the  map $w$ factors through $(\partial
f_s)\times\Sigma\times\hbox{point}$. When  $\ell<k-1$, $(\partial
f_s)\times\Sigma\times\hbox{point}$ has dimension  less than $k$, so
the differential is trivial. For $\ell=k-1$, this reduces  to the
claimed result. \hfill $\Box$
 
We remark that the above lemma is a valid computation of the
cohomology of  $G^K$ when $K$ is any connected $2$-complex.  To go up
to the next and final stage we need to analyze the action of the
fundamental group of $G^{M^{(2)}}$ on the cohomology of
$G^{M^{(3)}/M^{(2)}}$.  This will require Lemma \ref{surj} from the
beginning of this section. This  is the place in the cohomology
computation where we use the fact that $M$ is  orientable. Let us
review the situation for general fibrations first. If
$F\hookrightarrow E\to B$ is a fibration and $\gamma:[0, 1]\to B$
represents  an element of the fundamental group of $B$ one can define
a map,  $\Gamma:F\times [0, 1]\to B$ by $\Gamma(x,t)=\gamma(t)$. The
map,  $\Gamma_0:F\to B$ lifts to the inclusion, $F\to E$. By the
homotopy lifting  property, there is a lift, $\overline\Gamma:F\times
[0, 1]\to E$. The  restriction, $\overline\Gamma_1:F\to F$ induces a
map on the cohomology of  $F$. This is how the fundamental group of
the base of a fibration acts on the  cohomology of the fiber. The
following lemma shows that the action of the  fundamental group on the
cohomology of the fiber of our final fibration is  trivial.
\begin{lemma}
If $M$ is an orientable $3$-manifold and $G$ is a compact,
simply-connected  Lie group then the action of $\pi_1(G^{M^{(2)}})$ on
$H^*(G^{M^{(3)}/M^{(2)}})$ is trivial.
\end{lemma} 
{\it Proof:\, } Let $\widehat\gamma:SM^{(2)}\to G$ represent an
element of  $\pi_1(G^{M^{(2)}})$. By Lemma \ref{surj}, this extends to
a map,  $\widehat\zeta:SM^{(3)}\to G$. Now define
$\overline\Gamma:G^{M^{(3)}/M^{(2)}}\times [0, 1]\to G^M$ by
$\overline\Gamma(u, t)(x)=u([x])\widehat\zeta([x,t])$. The upper
triangle of  the following diagram commutes because
$\widehat\zeta([x,0])=1$. The lower  triangle commutes because,
$\widehat\zeta|_{SM^{(2)}}=\widehat\gamma$.
\begin{diagram}
G^{M^{(3)}/M^{(2)}}\times \{0\} & \rTo^{} &  G^M \\ \dTo^{} & \ruTo^{}
& \dTo_{} \\ G^{M^{(3)}/M^{(2)}}\times [0, 1] & \rTo_{} & G^{M^{(2)}}\\
\end{diagram}
Thus $\overline\Gamma$ is an appropriate lift. Since $u([x])=1$ for
$x\in M^{(2)}$, $\overline\Gamma_1$ is the identity map and the action
on the  cohomology of the fiber is trivial. \hfill $\Box$
 
We can now complete the proof of Theorem \ref{thm1}.  The cofibration
$M^{(2)}\to M^{(3)} \to M^{(3)}/ M^{(2)}$ leads to a  fibration,
$$
G^{M^{(3)}/ M^{(3)}}\to G^{M^{(3)}}\to G^{M^{(2)}}.
$$
By the previous lemma, the coefficients in the cohomology appearing in
the  second term of the Serre spectral sequence are untwisted. Using
Lemma  \ref{gm2}, we have,
$$
\begin{array}{rl}
E_2^{*,*}&=H^*(G^{M^{(2)}}; H^*(G^{M^{(3)}/M^{(2)}}))=
H^*(G^{M^{(2)}})\otimes H^*(G^{M^{(3)}/M^{(2)}}) \\ &\cong  {\mathbb
R}[\mu(\Sigma^1_r\otimes x_j),  \mu(\Sigma^2_s\otimes x_k),
\mu([M]\otimes x_\ell), \ell>0].
\end{array} 
$$
Repeating the argument from Lemma \ref{gm2} with computation
\ref{delta}, we see that all of the differentials of this spectral
sequence vanish. This completes our computation of the cohomology of
the  Skyrme configuration space. \hfill $\Box$

\subsection{The fundamental group of Faddeev-Hopf configuration spaces}
\label{subsecpc}

In this subsection we compute the fundamental group of the
Faddeev-Hopf  configuration space, $(S^2)^M$. Recall (Theorem
\ref{ponthm}) that the path components  $(S^2)^M_\varphi$ (where
$\varphi$ is any representative of the component) fall into families
labelled by $\varphi^*\mu_{S^2}\in H^2(M;\Z)$, where  $\mu_{S^2}$ is a
generator of $H^2(S^2;\Z)$, and that components within a given family
are labelled by $\alpha\in H^3(M;\Z)/2\varphi^*\mu_{S^2}\cp H^1(M;\Z)$.

To analyze the Faddeev-Hopf configuration space in more detail we will
further exploit its natural relationship with the classical
($G=\SU(2)=\Sp(1)$) Skyrme configuration space. These ideas were
concurrently introduced in \cite{AK2}.  We identify $S^2$ with the unit
purely imaginary quaternions, and $S^1$ with the unit complex
numbers. The quotient, Sp$(1)/S^1$ is homeomorphic to $S^2$, with an
explicit homeomorphism given by $[q]\mapsto qiq^*$.  Our main tool
will be the  map  
\beq\label{qdef} 
{\mathfrak q}:S^2\times S^1\to
\hbox{Sp}(1),\qquad {\mathfrak q}(x,\lambda)=q\lambda q^*,
\qquad\hbox{where} \ x=qiq^*.  
\eeq 
It is not difficult to verify the
following properties of ${\mathfrak q}$.
\begin{enumerate}
\item It is well defined and smooth.
\item ${\mathfrak q}(x,\lambda_1\lambda_2)= {\mathfrak q}(x,\lambda_1)
{\mathfrak q}(x,\lambda_2)$.
\item ${\mathfrak q}^{-1}(1)=S^2\times\{1\}$.
\item ${\mathfrak q}(x,\lambda)x({\mathfrak q}(x,\lambda))^*=x$.
\item ${\mathfrak q}(x,\cdot):S^1\to \{q|qxq^*=x\}$ is a
diffeomorphism.
\item deg$({\mathfrak q})=2$ with the standard ``outer normal first''
orientations. 
\end{enumerate}

For example writing $x=qiq^*$, the fourth property may be verified as
${\mathfrak q}(x,\lambda)x({\mathfrak q}(x,\lambda))^*=q\lambda
q^*qiq^*q\lambda^* q^*= x$.  We will let $\lambda^x$ denote the
inverse to ${\mathfrak q}(x,\cdot)$. We  will also use the related
maps,  $\rho: S^2\times\hbox{Sp}(1) \times S^1\to
S^2\times\hbox{Sp}(1)$ and  $f: S^2\times\hbox{Sp}(1)\to S^2\times
S^2$ defined by  $\rho(x,y,\lambda)=(x,y{\mathfrak q}(x,\lambda))$ and
$f(x, q)=(x, qxq^*)$.  
Properties (2) and (3) show that $\rho$ is a free right
action. Properties  (4) and (5) show that $f$ is a principal fibration
with action $\rho$.  As our first application of these maps we show
that the evaluation map is a fibration.
\begin{lemma}\label{s2tofree}
The evaluation map,
$$
\hbox{ev}_{p_0}:\hbox{FreeMap}(M, S^2) \to S^2,
$$
given by $\hbox{ev}_{p_0}(\varphi)=\varphi(p_0)$ is a fibration.
\end{lemma}
{\it Proof:\, } We just need to construct the diagonal map in the
following diagram.
\begin{diagram}
X\times \{0\} & \rTo^{\overline h} & \hbox{FreeMap}(M, S^2)   \\
\dTo^{i} & \ruTo^{\overline H} & \dTo_{\hbox{ev}_{p_0}} \\ X\times [0,
1] & \rTo_{H} & S^2 \\
\end{diagram}
If we define the horizontal maps in the following diagram by $\mu(x,
t)= (\overline h(x)(p_0), H(x,t))$ and  $\overline\nu(x)= (\overline
h(x)(p_0),1)$, then the existence of the diagonal map will follow
because $f$ is a fibration. 
\begin{diagram}
X\times \{0\} & \rTo^{\overline \nu} & S^2\times\hbox{Sp}(1)   \\
\dTo^{i} & \ruTo^{\overline \mu} & \dTo_{f} \\ X\times [0, 1] &
\rTo_{\mu} & S^2\times S^2\\
\end{diagram}
The desired map is just,  $\overline H(x, t)(p)= \overline\mu_2(x, t)
\overline h(x)(p)(\overline\mu_2(x, t))^*$ \hfill $\Box$

\noindent 
Clearly the fiber of this fibration is $(S^2)^M$.  Recall that we are
using $(S^2)^M_\varphi$ to denote the $\varphi$-component  of the
space of based maps. 

In \cite{AK2} these ideas were used to give a new proof of Pontrjagin's 
homotopy classification of maps from a $3$-manifold to $S^2$. The following 
lemma comes from that paper. A second proof of this lemma may be found in 
\cite{AK3}.

\begin{lemma}[Auckly-Kapitanski]
There exists a map $u:M\to \hbox{Sp}(1)$ such that $\psi:M\to S^2$ and 
$\varphi:M\to S^2$ are related by $\psi=u\varphi u^*$ if and only if 
$\psi^*\mu_{S^2}=\varphi^*\mu_{S^2}$.
\end{lemma}

Theorem \ref{fhhom} follows directly from this lemma. Assuming 
$\psi^*\mu_{S^2}=\varphi^*\mu_{S^2}$, define a map $F: (S^2)^M_\varphi 
\to (S^2)^M_\psi$ by $F(\xi)=u\xi u^*$. This is clearly well defined 
because any map homotopic to $\varphi$ will be mapped to a map homotopic 
to $\psi$ under $F$. There is a well-defined inverse given by 
$F^{-1}(\zeta)=u^*\zeta u$.

We have a fibration relating the identity
component of the Skyrme configuration space to any component of the
Faddeev-Hopf configuration space.

\begin{lemma}\label{Fhseq}
The map induced by $f$,
$$
\{\varphi\}\times \hbox{Sp}(1)^M_0
\stackrel{f^*}{\rightarrow}\{\varphi\}\times (S^2)^M_\varphi
$$
is a fibration.
\end{lemma}
{\it Proof:\, } Once again we just need to construct the diagonal map
in a diagram.
\begin{diagram}
X\times \{0\} & \rTo^{\overline h} & \hbox{Sp}(1)^M_0   \\ \dTo^{i} &
\ruTo^{\overline H} & \dTo_{f_*} \\ X\times [0, 1] & \rTo_{H} &
(S^2)^M_\varphi\\
\end{diagram}
So once again we consider a second diagram.
\begin{diagram}
M\times X\times \{0\} & \rTo^{\overline \nu} & S^2\times\hbox{Sp}(1)
\\ \dTo^{i} & \ruTo^{\overline \mu} & \dTo_{f} \\ M\times X\times [0,
1] & \rTo_{\mu} & S^2\times S^2\\
\end{diagram}
Here the horizontal maps are given by $\overline\nu(p, x)=(\varphi(p),
\overline h(x)(p))$ and $\mu(p, x, t)=(\varphi(p), H(x, t)(p))$. The
diagonal  lift exists because $f$ is a fibration. We need to use
property (5) of  ${\mathfrak q}$ to adjust the base points. Let $x_0$
be the basepoint of  $S^2$ and define the desired lift by
$$
\overline H(x, t)(p)=\overline\mu_2(p, x, t) {\mathfrak
q}(\varphi(p),\lambda^{x_0} (\overline\mu_2(p, x, t)^{-1}).
$$
This completes the proof. \hfill $\Box$
 
By property (5) of ${\mathfrak q}$, we see that any element of the
fiber of  the above fibration may be written in the form  ${\mathfrak
q}(\varphi, \lambda)$ for some map $\lambda:M\to S^1$. Since
${\mathfrak q}(\varphi, \lambda)$ is null homotopic, its degree must
be  zero. By property (6) of ${\mathfrak q}$, this implies that
$\lambda^*\mu_{S^1}$ must be in the kernel of the cup product
$2\varphi^*\mu_{S^2}\cp$. Conversely, given any map $\lambda$ with
$\lambda^*\mu_{S^1}\in\hbox{ker}(2\varphi^*\mu_{S^2}\cp)$ we get an
element of the fiber. Recall that the components of the space of maps
from  $M$ to $S^1$ correspond to $H^1(M;{\mathbb Z})$ and each
component is  homeomorphic to the identity component which is
homeomorphic to  ${\mathbb R}^M$ which is contractible. It follows
that up to homotopy  Sp$(1)^M_0$ is a regular
ker$(2\varphi^*\mu_{S^2}\cp)$ cover of  $(S^2)^M_\varphi$ (the fiber
is homotopy equivalent to ker$(2\varphi^*\mu_{S^2}\cp)$).   The homotopy
sequence of the fibration then gives us the following sequence:
\begin{equation}\label{fhseq}
0\to \pi_1(\hbox{Sp}(1)^M)\to \pi_1((S^2)^M_\varphi)\to
\hbox{ker}(2\varphi^*\mu_{S^2}\cp)\to 0.
\end{equation}
Since we do not already know that $\pi_1((S^2)^M_\varphi)$ is abelian,
we  not only need to show that the sequence splits, we also need to
show that  the image of the splitting commutes with the image of
$\pi_1(\hbox{Sp}(1)^M)$.  This is the content of the following
lemma. This lemma will complete the  proof of Theorem \ref{thm2}.

\begin{lemma}\label{Fhsplit} The sequence (\ref{fhseq}) splits and the 
image of the splitting
commutes with the image of $\pi_1(\hbox{Sp}(1)^M)$.
\end{lemma}
{\it Proof:\, } Given $\theta\in\hbox{ker}(2\varphi^*\mu_{S^2})$
define a corresponding map  $\lambda_\theta:M\to S^1$ in the usual
way,  $\lambda_\theta(p)=e^{\int^p_{p_0}\theta}$. This induces a map,
$q_\theta:M\to \hbox{Sp}(1)$ by  $q_\theta(p)={\mathfrak
q}(\varphi(p),\lambda_\theta(p))$.  We compute the degree as follows:
$$
\begin{array}{rl}
\hbox{deg}(q_\theta)= & \int_M q_\theta^*\mu_{\hbox{Sp}(1)} \\ = &
2\int_M \varphi^*\mu_{S^2}\wedge\lambda^*_\theta\mu_{S^1} = 2\int_M
\varphi^*\mu_{S^2}\wedge\lambda^*_\theta = 0.
\end{array}
$$
It follows that there is a homotopy, $\overline H_\theta$ with
$\overline H_\theta(0)=1$ and $\overline H_\theta(1)=q_\theta$. Define
the  splitting by sending $\theta$ to
$H_\theta\in\pi_1((S^2)_\varphi)$ given by
$H_\theta(t)(p)=f(\varphi(p),\overline H_\theta(t)(p))$. To see that
the  image of this splitting commutes with the image of
$\pi_1(\hbox{Sp}(1)^M)$,  let $\gamma:[0,1]\to \hbox{Sp}(1)^M$ be a
loop and define maps $\delta_1$ and  $\delta_2$ by
$\delta_1(t,s)=\frac{2t}{s+1}$ for $t\le \frac12 (s+1)$,
$\delta_1(t,s)=1$ otherwise and $\delta_2(t,s)=1-\delta_1(1-t,s)$.  We
see that $f(\varphi,(\gamma\circ\delta_1)\cdot (\overline
H_\theta\circ\delta_2))$ is a homotopy between  $f(\varphi,
\gamma)*f(\varphi,\overline H_\theta)$ and  $f(\varphi,
\gamma\cdot\overline H_\theta)$. Likewise,
$f(\varphi,(\gamma\circ\delta_2)\cdot(\overline
H_\theta\circ\delta_1))$  is a homotopy between $f(\varphi,\overline
H_\theta)* f(\varphi, \gamma)$ and  $f(\varphi, \gamma\cdot\overline
H_\theta)$.   \hfill $\Box$

To prove Theorem \ref{freefun}, notice that we have a left $S^1$
action on  $(S^2)^M_\varphi$ given by $z\cdot\psi:=z\psi z^*$. We
claim that the fibration,  $ \hbox{FreeMap}(M, S^2)_\varphi \to S^2, $
is just the fiber bundle with associated principal bundle Sp$(1)\to
S^2$ and fiber $(S^2)^M_\varphi$. In fact, the map
Sp$(1)\times_{S^1}(S^2)^M_\varphi \to \free{S^2}{M}$ given by
$[q,\psi]\mapsto q\psi q^*$ is the desired isomorphism. Now consider
the homotopy exact sequence of the fibration, $ \hbox{FreeMap}(M,
S^2)_\varphi \to S^2, $
$$
\to\pi_2(S^2)\to \pi_1((S^2)^M_\varphi) \to
\pi_1(\free{S^2}{M}_\varphi) \to 0.
$$
It follows that $\pi_1(\free{S^2}{M}_\varphi)$ is just the quotient of
$\pi_1((S^2)^M_\varphi)$ by the image of $\pi_2(S^2)$. The next lemma
identifies this image, to complete the proof of Theorem \ref{freefun}.

\begin{lemma}
The image of the map from $\pi_2$ is the subgroup of $H^2(M;\Z)<
\pi_1((S^2)^M_\varphi)$ generated by $2\varphi^*\mu_{S^2}$.
\end{lemma}
{\it Proof:\, } Recall that the map from $\pi_2$ of the base to
$\pi_1$ of the fiber is defined by taking a map of a disk into the base to
the restriction to the boundary of a lift of the disk to the total
space. Since the boundary of the disk maps to the base point, the
restriction to the boundary of the lift lies in the fiber. The
homotopy exact sequence of the fibration, Sp$(1)\to S^2$ implies that
the disk representing a generator of $\pi_2(S^2)$ lifts to a disk with
boundary generating the fundamental group of the fiber $S^1$. This
lift, say $\widehat\gamma$ to Sp$(1)$ gives a lift $D^2\to
\hbox{Sp}(1)\times_{S^1}(S^2)^M_\varphi \cong\free{M}{S^2}$ defined by
$z\mapsto [\widehat\gamma(z),1]$. Restricted to the boundary, this map
is just $z\mapsto [z,\varphi]=[1,z\varphi z^*]$. It follows that the
image of $\pi_2(S^2)$ is just the subgroup generated by the loop
$z\varphi z^*$. We now just have to trace this loop through the proof
of the isomorphism, 
$$
\pi_1((S^2)^M_\varphi) \cong {\mathbb Z}_2\oplus H^2(M;{\mathbb
Z})\oplus  \hbox{\rm ker}(2\varphi^*\mu_{S^2}\cp).
$$
The projection to $\hbox{\rm ker}(2\varphi^*\mu_{S^2}\cp)$ was defined
 by taking a lift of each map in the $1$-parameter family representing
 the loop in $\pi_1((S^2)^M_\varphi)$ to Sp$(1)^M_0$ and comparing the
 maps at the beginning and end. In our case the entire path
 consistently lifts to  the path $\gamma_\varphi:S^1\to
 \hbox{Sp}(1)^M_0$ given by $\gamma_\varphi(z)=z{\mathfrak
 q}(\varphi,z^*)$. It follows that the component in $\hbox{\rm
 ker}(2\varphi^*\mu_{S^2}\cp)$ is zero. A loop such as
 $\gamma_\varphi$ naturally defines a map, $\bar\gamma_\varphi:M\times
 S^1\to \hbox{Sp}(1)$. The image of our loop in $H^2(M;\Z)$ is just
 $\bar\gamma_\varphi^*\mu_{{\rm Sp}(1)}/[\hbox{pt}\times S^1]$. In
 notation reminiscent of differential forms this would be
 $\int_{{\rm pt}\times S^1 }\bar\gamma_\varphi^*
 \mu_{{\rm Sp}(1)}$. In order to evaluate this, we write
 $\bar\gamma_\varphi$ as the composition of the map
 $(\varphi,\hbox{id}_{S^1}):M\times S^1\to S^2\times S^1$ and the map
 $\tilde{\mathfrak q}:S^2\times S^1\to \hbox{Sp}(1)$ given by
 $\tilde{\mathfrak q}(x,z)=z{\mathfrak q}(x,z^*)$. This latter map is
 then expressed as the composition of $({\mathfrak q},
 \hbox{pr}_2^*):S^2\times S^1 \to \hbox{Sp}(1)\times S^1$ and the map
 $\hbox{Sp}(1)\times S^1\to \hbox{Sp}(1)$ given by $(u,\lambda)\mapsto
 \lambda u$. The form $\mu_{{\rm Sp}(1)}$ pulls back to
 $\mu_{{\rm Sp}(1)}\cup 1$ under the first map, and this pulls back
 to $2\mu_{S^2}\cup\mu_{S^1}$ under the first factor of
 $\tilde{\mathfrak q}$ since ${\mathfrak q}$ has degree two. In
 particular $\tilde{\mathfrak q}$ has degree two as well. We can now
 complete this computation to see that our loop projects to
 $2\varphi^*\mu_{S^2}$ in $H^2(M;\Z)$. To complete the proof, we need
 to compute the projection of our loop in the $\Z_2$-factor.  The
 projection to $\Z_2$ is defined by multiplying the inverse of our map
 by the image of $2\varphi^*\mu_{S^2}$ under the splitting
 $H^2\to\pi_1$ and taking the framing of the inverse image of a
 regular point. The equivalence classes of framings may be identified
 with $\Z_2$ since the inverse image is homologically
 trivial. Alternatively we may compare the framing coming from our map
 to the framing of the map coming from the splitting. The image under
 the splitting of $2\varphi^*\mu_{S^2}$ is, of course, just two times
 the image of $\varphi^*\mu_{S^2}$ under the splitting. The inverse
 image coming from our map is just two copies of the inverse image of
 a frame under the map $(\varphi,\hbox{id}_{S^1}):M\times S^1\to
 S^2\times S^1$. This means that the projection is even, so zero in
 $\Z_2$.   \hfill $\Box$   

\subsection{The cohomology of Faddeev-Hopf configuration spaces}
\label{subsecpd}

In order to compute the cohomology of $(S^2)^M_\varphi$ we will use
the fibration, Sp$(1)^M_0\to (S^2)^M_\varphi$. The fiber of this
fibration is just $\disju_{\alpha\in K} (S^1)^M_\alpha$ where
$K=\hbox{ker}(2\varphi^*\mu_{S^2}\cup)$. Up to homotopy, the fiber is
just $K$. In fact we can assume that the fiber is exactly $K$ if we
first take the quotient by $(S^1)^M_0$ (which is contractible by
Reduction \ref{r6}). It is slightly tricky to use a spectral sequence
to compute the cohomology of the base of a fibration, so we will use
equivariant cohomology to recast the problem.  Recall that any Lie
group acts properly on a contractible space called the total space of
the universal bundle. In our case, this space is denoted $EK$. The
quotient of this by $K$ is the classifying space $BK$. Let $X$ be a
$K$ space (i.e. a space with a $K$ action) and consider the space
$X_K:=EK\times_K X$. We will be interested in the situation when
$X=\hbox{Sp}(1)^M_0$ (really this divided by $(S^1)^M_0$ but this has
the same homotopy type).  The cohomology of the space $X_K$ is called
the equivariant cohomology of $X$. It is denoted by $H^*_K(X)$. When
the $K$ action on $X$ is free and proper (as it is in our case), we
have a fibration $X_K\to X/K$ obtained by ignoring the $EK$ component
in the definition of $X_K$. The fiber of this fibration is just $EK$
which is contractible, so the spectral sequence of the fibration
implies that the cohomology of $X/K$ is isomorphic to the equivariant
cohomology of $X$. By ignoring the $X$ component in the definition of
$X_K$ we obtain a fibration $X_K\to BK$ which may be used to compute
the equivariant cohomology of $X$. 

Since $H^1(M;\Z)$ is a free abelian group, the kernel $K$ is as
well. It follows that we may take $EK$ to be $\R^n$ with $n$ equal to
the rank of $K$ and with $K$ acting by translations. It follows that
$BK$ is just an $n$-torus, and we have a spectral sequence with $E_2$
term, $E_2^{p,q}=H^p(T^n;\tilde{ H^q(\hbox{Sp}(1)^M_0) })$ converging
to the cohomology of $(S^2)^M_\varphi$. Clearly the fundamental group
of $T^n$ is just $K$. To compute the action of $K$ on
$H^*(\hbox{Sp}(1)^M_0)$, let $\lambda:M\to S^1$ satisfy
$\lambda^*\mu_{S^1}\in K$, and $\mu(\Sigma\otimes x)\in
H^q(\hbox{Sp}(1)^M_0)$ with $\sigma:\Sigma\to M$. Let
$u:\Delta^q\to\hbox{Sp}(1)^M_0$ be a singular $q$-simplex and let
$m:\hbox{Sp}(1)\times\hbox{Sp}(1)\to\hbox{Sp}(1)$ be the
multiplication. Then we have,
$$
\begin{array}{rl}
(\lambda^*\mu_{S^1}\cdot\mu(\Sigma\otimes x))(u) &=
\int_{\Sigma\times\Delta^q} m(\widehat u,{\mathfrak
q}(\varphi,\lambda) \circ(\sigma,1))^*x \\ &=
\int_{\Sigma\times\Delta^q} (\sigma,1)^* {\mathfrak
q}(\varphi,\lambda)^*L_{\widehat u}^*x+R_{{\mathfrak
q}(\varphi,\lambda) \circ(\sigma,1)}^*\widehat u^*x  \\
&=\int_{\Sigma\times\Delta^q}\widehat u^*x = \mu(\Sigma\otimes x)(u).
\end{array}
$$
Thus the fundamental group of the base acts trivially on the
cohomology of the fiber. Because this fibration has an associated
principal fibration with discrete group, all of the higher
differentials vanish, and we obtain Theorem \ref{thm2co}.

Theorem \ref{freeco} will follow from considerations of a general
fiber bundle with structure group $S^1$ and one computation. Let $P\to
X$ be a principal $S^1$ bundle with simply-connected base and let
$\tau:S^1\times F\to F$ be a left action. The Serre spectral sequence
of the fibration $E=P\times_{S^1}F\to X$ has $E_2^{p,q}=H^p(X;
{H^q(F;\R)})$ and second differential $d_2\omega=c_1(P)\cup
\tau^*\omega/[S^1]$. In our case, the principal bundle is Sp$(1)\to
S^2$. It follows immediately that the coefficient system in the $E_2$
term of the Serre spectral sequence is untwisted, and that the only
non-trivial differential is the $d_2$ differential. In this case, the
first Chern class is $\mu_{S^2}$. The action that we consider is the
map $\tau:S^1\times (S^2)^M_\varphi$ given by $\tau(z,u)=zuz^*$. In
fact, we only need to consider the effect of this action on terms
coming from Sp$(1)^M_0$. This is because the action is trivial on the
classes coming from $BK$. This can be seen by considering a map from
$(S^2)^M_\varphi$ to $BK$. However, the easiest way to see this is first to
compute the cohomology of the fiber bundle with fiber
Sp$(1)^M_0$, and then recognize that, up to homotopy, the total space of
this bundle is a regular $K$-cover of $\free{M}{S^2}_\varphi$. Either
way, we need to compute the second differential coming from the action,
$\tau_0:S^1\times \hbox{Sp}(1)^M_0\to\hbox{Sp}(1)^M_0$ given by
$\tau_0(z,u)=\tilde{\mathfrak q}(\varphi,z)u$. Let $u:F\to
\hbox{Sp}(1)^M_0$ be a singular chain and compute 
$$
\left(\tau_0^*\mu(\Sigma\otimes x)/[S^1]\right)(u) =
\int_{S^1\times\Sigma\times F} \left(m\circ(\tilde{\mathfrak
q}(\varphi\circ\sigma,\hbox{pr}_{S^1}),\widehat
u\circ(\sigma,\hbox{pr}_F)\right)^*x.
$$
Here $m:\hbox{Sp}(1)\times\hbox{Sp}(1)\to\hbox{Sp}(1)$ is
multiplication, and the rest of the maps are as in the definition of
$\mu(\Sigma\otimes x)$ in line \ref{mudef}. This vanishes for
dimensional reasons when $\Sigma$ is a $1$-cycle ($\varphi\circ\sigma$
would push it forward to a $1$-cycle in $S^2$). When $\Sigma$ is a
$2$-cycle, we use the product rule and the fact that $\tilde{\mathfrak
q}:S^2\times S^1\to\hbox{Sp}(1)$ has degree two to conclude that
$\left(\tau_0^*\mu(\Sigma\otimes
x)/[S^1]\right)=2\varphi^*\mu_{S^2}[\Sigma]$. This completes the proof
of our last theorem.

\end{document}